\documentclass[11pt,english]{article}

\usepackage{bm}
\usepackage[T1]{fontenc}
\usepackage[utf8]{inputenc}
\usepackage{lmodern}
\usepackage{amsmath}
\usepackage{amssymb}
\usepackage{cancel}
\usepackage[english]{babel}
\usepackage{physics}
\usepackage[toc,page]{appendix}
\usepackage{comment}
\usepackage{cite}
\usepackage{authblk}
\usepackage{graphicx} 
\usepackage{float}
\usepackage[left=25mm,top=26mm,right=20mm,bottom=25mm]{geometry}
\usepackage{xcolor}

\usepackage{caption}
\usepackage{subcaption}
\usepackage{physics}
\usepackage[export]{adjustbox}
\usepackage{caption}

\usepackage{epstopdf} 
\usepackage{comment}
\usepackage{tabularx} 
\usepackage{booktabs} 
\usepackage{psfrag} 

\usepackage{eurosym} 

\usepackage{xcolor} 
\usepackage{empheq} 
\definecolor{Hamzacolor}{rgb}{1000,00,100}
\definecolor{linkcolor}{rgb}{0,0,0.6} 
\usepackage[ 
    pdftex,
    colorlinks=true,
    pdfstartview=FitV,
    linkcolor=linkcolor,
    citecolor=linkcolor,
    urlcolor=linkcolor,
    hyperindex=true,
    hyperfigures=false,
    backref=page,            
]{hyperref}
\renewcommand*{\backref}[1]{}
\renewcommand*{\backrefalt}[4]{%
  \ifcase #1 %
    (Not cited.)%
  \or
    (cit. on p.~#2)%
  \else
    (cit. on pp.~#2)%
  \fi
}

\usepackage{fancyhdr} 
\usepackage{pgfplots} 
\usepgfplotslibrary{groupplots}

\newcommand{\Q}{Q}

\newcommand\cA{\mathcal{A}}
\newcommand\cB{\mathcal{B}}

\newcommand\nth{{p}}

\newcommand\Ftwo{{F}}
\newcommand\X{{\bar X}}

\newcommand\p{P}
\newcommand\bb{\tau}

\newcommand{\tb}{b}

\numberwithin{equation}{section}
\numberwithin{figure}{section}
\numberwithin{table}{section}
\newcolumntype{Y}{>{\centering\arraybackslash}X}
\pgfplotsset{compat=1.6}
\pgfplotsset{/pgf/number format/use comma}

\author[a]{Hamza Boumaza}
\author[b]{Christos Charmousis}
\author[c]{David Langlois}
\author[c,d]{Etienne Ligout}
\title{Neutron stars with primary scalar hair}
\affil[a]{\textit{Laboratoire de Physique des Particules et Physique Statistique} (LPPPS), \textit{Ecole Normale Supérieure-Kouba}, B.P. 92, Vieux Kouba, 16050 Algiers, Algeria}
\affil[b]{\textit{Université Paris-Saclay}, CNRS/IN2P3, IJCLab, 91405 Orsay, France}
\affil[c]{\textit{Université Paris Cité}, CNRS, \textit{Astroparticule et Cosmologie}, F-75013 Paris, France}
\affil[d]{\emph{Sorbonne Universit{\'e}}, CNRS, \emph{Institut d'Astrophysique de Paris}, $\mathcal{G}\mathbb{R}\varepsilon{\mathbb{C}}\mathcal{O}$, 75014 Paris, France}
\begin{document}

\maketitle
\begin{center}
{\bf Abstract}
\end{center}
  We investigate static and spherically symmetric neutron star solutions endowed with primary scalar hair in a subfamily of Degenerate-Higher-Order-Scalar-Tensor (DHOST) theories of gravity. 
By solving the modified Tolman-Oppenheimer-Volkoff (TOV)  equations, we construct equilibrium configurations for polytropic and realistic equations of state and analyse the impact of the scalar hair on the stellar structure. We examine the resulting metric and  scalar field profiles as well as the mass-radius relation, showing deviations from the predictions of General Relativity (GR). 
Positive scalar charges lead to more compact stars than in GR and, above a critical threshold, to singularities. Observations could therefore put stringent constraints on the parameters characterising the beyond-GR effects in these theories and their potential scalar hair.

\newpage
\tableofcontents

\section{Introduction}
Strong gravity manifests itself in compact objects, such as black holes and neutron stars, making them ideal arenas for testing General Relativity (GR) and exploring alternative theories of gravity. In recent years 
a wealth of observational data on compact objects has been collected by
gravitational-wave (GW) detectors \cite{LIGOScientific:2018mvr,LIGOScientific:2026wfs},  networks of radio-telescopes like the Event Horizon Telescope \cite{ EventHorizonTelescope:2020qrl} and X-ray telescopes such as NICER\cite{Raaijmakers:2019dks}. Observational data will be greatly enhanced in the near future with far better precision as well as novel instruments such as the Einstein Telescope \cite{ET:2025xjr} or the Laser Interferometer Space Antenna (LISA) \cite{LISA:2017pwj}.

While current data remain compatible with GR within observational uncertainties, it is crucial to anticipate potential effects beyond GR. This preparation is essential both for interpreting the upcoming plethora of data and for exploring alternative scenarios motivated by open questions or existing tensions.
Beyond direct consistency checks of GR \cite{LIGOScientific:2026qni},  important questions remain to be answered, such as the 
equation of state (EoS) of high-density nuclear matter,
the maximum mass of neutron stars, their maximum rotation, etc. Furthermore, certain tensions with observations exist. For example recent observations of compact binaries such as GW190814 \cite{LIGOScientific:2020zkf}, GW200210 or  GW230529 \cite{LIGOScientific:2024elc} give primary or secondary compact objects with a mass well within the mass gap interval between $2$ and $5$ $M_\odot$ predicted in astrophysical processes in GR. In the case of GW190814, the secondary component  of the binary has a mass of $2.59_{-0.09}^{+0.08}~M_\odot$. Within GR,  this could only  be explained by a neutron star with an unexpectedly stiff (or exotic) EoS, an unsually rapidly rotating neutron star  or a black hole with a surprisingly  small mass (see for example \cite{Huang:2020cab,Bombaci:2020vgw,Zhou:2020xan,Most:2020bba,Kanakis-Pegios:2020kzp,Nathanail:2021tay}).

Many theories of modified gravity are based on the inclusion of a scalar field in the gravitational sector, in addition to the usual metric field. The most general scalar-tensor theories propagating a single scalar degree of freedom are known as Degenerate-Higher-Order-Scalar-Tensor (DHOST) theories\footnote{Despite the presence of second-order derivatives of the scalar field in their Lagrangian, these theories avoid the presence of a ghost-like extra degree of freedom due to degeneracy relations satisfied by their Lagrangian functions.}~\cite{Langlois:2015cwa,Langlois:2015skt,BenAchour:2016fzp} (see \cite{Langlois:2018dxi,Kobayashi:2019hrl} for reviews), which include  well studied theories such as the traditional scalar-tensor theories,  Horndeski theories and Einstein-scalar-Gauss-Bonnet theories. 
Recently, new static black hole (BH) solutions --- characterised by  primary hair ---  have been discovered \cite{Bakopoulos:2023fmv} (see also \cite{Baake:2023zsq,Bakopoulos:2023sdm}) within the framework of DHOST theories. These BH solutions represent a one-parameter deformation of the traditional Schwarzschild metric, together with a non-trivial scalar field profile. The primary scalar hair of these solutions is due to a global shift symmetry of the action which is translated in a linear time dependence for the scalar field \cite{Babichev:2013cya}. This not only gives an independent scalar hair but also renders the scalar field regular at the event horizon of the black hole. 

The extension of these static black holes to slow rotation was recently achieved in \cite{Candan:2025fbl}, while theoretical studies on their intriguing thermodynamic properties were initiated in \cite{Bakopoulos:2024ogt, Bakopoulos:2024zke, Erices:2024lci}. More phenomenological aspects such as strong lensing and shadow analyses, were recently explored in \cite{Fathi:2025uwa, Nozari:2026wjo}. Linear perturbations of these solutions, in particular quasinormal modes,  were investigated in \cite{Sirera:2024ghv, Antoniou:2025bvg, Montagnon:2025zfl, Lahoz:2025mwt, Konoplya:2026cus},  while  axial  
and radial 
perturbations around  these backgrounds were studied in 
\cite{Charmousis:2025xug,Charmousis:2026bhe}.

In this work, we extend the study of the same subfamily of DHOST theories to relativistic stars --- typically neutron stars. While the exterior of these stars matches  that of the BH solutions reported in  \cite{Bakopoulos:2023fmv},  the presence of matter modifies their interior. To compute the star's interior profile, we derive  generalised Tolman-Oppenheimer-Volkoff (TOV) equations that incorporate the effects of the gravitational scalar field. Interestingly, we find that these  modified gravity effects are encapsulated in a radially-dependent effective perfect fluid  with equation of state  $P_{\rm eff}=-\rho_{\rm eff}$. As a consequence, for the same central density of baryonic matter, the mass and the radius of the star differ from their GR counterparts. For a given DHOST theory, all these modifications depend on a theory-dependent length parameter and on the effective scalar charge of the  star. 
By integrating the modified TOV equations, we obtain numerical solutions that represent one-parameter  deformations of  GR neutron stars, characterised solely by their central density and EoS.

Starting with the well-known scalarisation mechanism proposed in \cite{Damour:1993hw}, numerous  investigations of relativistic stars within different distinct subfamilies of DHOST theories have been reported in the literature (see e.g. \cite{Cisterna:2015yla,Cisterna:2016vdx,Babichev:2016jom, Sakstein:2016oel,Babichev:2016rlq,Lehebel:2017fag,Chagoya:2018lmv,Kobayashi:2018xvr,Ogawa:2019gjc,Ventagli:2021ubn,Minamitsuji:2022tze,Saavedra:2024fzy,Diedrichs:2025vhv,Kobayashi:2025bdh}). For a particular class of Horndeski theories \cite{Lu:2020iav,Hennigar:2020lsl,Fernandes:2022zrq} it was found in \cite{Charmousis:2021npl} that neutron stars could acquire a larger mass than in GR (such as that observed in certain GW binaries \cite{LIGOScientific:2020zkf, LIGOScientific:2024elc}) without requiring a stiffer EoS. Our study   also exhibits this property, but with a key difference: in our model,  a more massive star  requires a  positive scalar charge, whereas in \cite{Charmousis:2021npl}, it required a positive coupling constant. In other words, the fixed coupling constant in the earlier theory is replaced here by  a free parameter --- the primary hair.

The outline of our paper is the following. In the next section, we introduce  the  subfamily of DHOST  theories we will focus on  and   recall the recently discovered BH  solutions with primary scalar hair in this framework.  We then derive, in 
Section \ref{section_EoM}, the modified TOV equations governing  a static and spherically symmetric configuration. In 
Section \ref{section_numerics}, we solve numerically the system of equations for a polytropic EoS and a few phenomenological EoS. 
We then give some conclusions and perspectives in the final section. Two appendices are added to provide some technical details.

\section{Model and previous solutions}
In this section, we present DHOST theories and review the static BH solutions  with primary hair that will describe the exterior of  our neutron star solutions. These black hole solutions have been discovered in \cite{Bakopoulos:2023fmv} and, soon after,  extended in \cite{Baake:2023zsq,Bakopoulos:2023sdm}. Their axial and radial perturbations and several other properties were recently analysed in \cite{Charmousis:2025xug,Charmousis:2026bhe}, whose notation we adopt. We will in particular focus on the homogeneous\footnote{By homogeneous metric, we mean one such that $g_{tt}=-1/g_{rr}$, as in Schwarzschild.} versions of these BH spacetimes since ---  as  explicitly shown in \cite{Bakopoulos:2023sdm,Charmousis:2025xug} --- any non homogeneous solution belongs to an equivalence class which is connected to a homogeneous solution via a disformal transformation~\cite{BenAchour:2016cay}.

\subsection{DHOST theories, general Lagrangian}

We consider DHOST theories whose action contains up to quadratic terms in  second derivatives of the scalar field. These theories are described by the action \cite{Langlois:2015cwa}
\begin{align}
\label{DHOSTaction}
S\left[g_{\mu \nu}, \phi\right]= \frac{1}{2\kappa}\int \mathrm{d}^4 x \sqrt{-g} \left( P(X, \phi)+ Q(X, \phi) \square \phi+\Ftwo(X, \phi) R +  \sum_{i=1}^5 A_i(X, \phi) L_i^{(2)} \right)   
\end{align}
where $\kappa\equiv 8\pi G$ and the kinetic density is given by\footnote{Be aware that one often encounters the alternative convention $X\equiv \partial_{\mu}\phi\, \partial^\mu \phi$.}
\begin{equation}
    X\equiv  -\frac{1}{2} \partial_\mu \phi \, \partial^\mu \phi\,.
\end{equation} 
The theory functions $P$, $Q$, $\Ftwo$ and $A_i$ depend on $\phi$ and $X$, while the elementary quadratic Lagrangians read~\cite{Langlois:2015cwa} 
\begin{eqnarray*}
&L_{1}^{(2)}=
\phi^{\mu\nu}\phi_{\mu\nu}\,, \qquad L_{2}^{(2)}=\left( \square\phi\right)^{2}\,, \qquad L_{3}^{(2)}= (\square\phi)\phi^{\mu}\phi_{\mu\nu}\phi^\nu\,,
\qquad \\
  & L_{4}^{(2)}=\phi^\lambda\phi_{\lambda\mu}\phi^{\mu\nu}\phi_\nu \,, \qquad \qquad
  L_{5}^{(2)}=\left(\phi^{\mu}\phi_{\mu\nu}\phi^\nu\right)^{2}\,,
  \qquad
\end{eqnarray*}
with the  notation $\phi_\mu\equiv \nabla_{\mu}\phi$ and $\phi_{\mu\nu}\equiv \nabla_{\nu}\nabla_{\mu}\phi$.
The functions $\Ftwo$ and $A_i$ satisfy three degeneracy conditions, ensuring that  the resulting theory  contains only a single scalar degree of freedom and thus avoids Ostrogradski instabilities.  

The theories discussed in this paper all belong to a sub-class of
DHOST theories, known as Beyond Horndeski theories~\cite{Gleyzes:2014dya,Gleyzes:2014qga}, characterised by the conditions
\begin{eqnarray}
\label{Ai_GLPV}
A_2=-A_1   \,, \qquad A_4=-A_3= \frac{A_1+\Ftwo_{,X}}{X} \,, \qquad A_5=0\,,
\end{eqnarray}
which ensure that the degeneracy conditions are verified. This sub-class itself contains the Horndeski theories~\cite{Horndeski:1974wa}, which satisfy
\begin{eqnarray}
\label{Ai_Horndeski}
A_2=-A_1= \Ftwo_{,X}  \,, \qquad A_3=A_4=A_5=0\,.
\end{eqnarray}
From now on, we restrict our attention to Lagrangians in which $Q=0$ and the remaining independent functions $P$, $\Ftwo$ and $A_1$ depend only on $X$. This implies that  the theory possesses shift ($\phi \rightarrow \phi \, + $ const)  and parity  ($\phi \rightarrow -\phi$) global symmetries.

\subsection{Black hole solutions}

\medskip
We consider a general ansatz for static and  spherically symmetric metrics, given by
\begin{align}
\label{background_metric}
\mathrm{d} s^2 =-\cA(r) \mathrm{d} t^2+\frac{\mathrm{d} r^2}{\cB(r)} +r^2 \mathrm{d} \Omega^2\,, \qquad \mathrm{~d} \Omega^2=\mathrm{d} \theta^2+\sin^2\!\theta\, \mathrm{d} \varphi^2\,.
\end{align}
Due to the  global shift symmetry of the theory, the scalar field can acquire a linear time dependence,
\begin{align}
\label{scalaransatz}
    \phi(t,r)=q t+\psi(r)\,,
\end{align}
where $q$ is an integration constant. The presence of this integration constant
circumvents the  requirements of no-hair theorems\cite{Hui:2012qt,Babichev:2016rlq} and gives rise to an independent primary hair. Indeed, the field equations in vacuum reduce to three simple independent equations~\cite{Bakopoulos:2023fmv}:
\begin{equation}
\label{easy}
 \frac{\cA}{\cB}=\frac{\gamma^2}{Z^2},
\end{equation}
\begin{equation}
\label{easy2}
    r^2(P Z)_{X}+2(\Ftwo Z)_X \left(1-\frac{q^2 \gamma^2}{2 Z^2 X}\right)=0,
\end{equation}
\begin{equation}
\label{easy3}
    2\gamma^2 \left(\cA r-\frac{q^2 r}{2 X}\right)'=-r^2 P Z-2\Ftwo Z\left(1-\frac{q^2 \gamma^2}{2 Z^2 X}\right)+\frac{q^2\gamma^2X' r}{ZX^2}\left( 2X\Ftwo_{X}-\Ftwo\right),
\end{equation}
where\footnote{In the Beyond Horndeski notation, as used in the original papers \cite{Bakopoulos:2023fmv,Baake:2023zsq,Bakopoulos:2023sdm}, \eqref{Z} becomes $Z=-G_4+2 X G_{4X}+4 X^2 F_4$.} 
\begin{equation}
\label{Z}
    Z\equiv -\Ftwo-2 X A_1 \,,
\end{equation}
a subscript $X$ denotes a derivative  with respect to $X$ and  a prime  a derivative with respect to the radial coordinate $r$. The three equations include an integration constant $\gamma$ which we will fix to unity without loss of generality. Hence \eqref{easy} dictates that homogeneous black holes, i.e. such that  $\cA=\cB$, are only possible\footnote{Since the GR limit corresponds to $F=1$, we choose the negative sign for $Z$, consistent  with its definition \eqref{Z}.} for $Z=-1$  while any other non-homogeneous solution will be parametrised by a non trivial $Z$. In principle,  any non-homogeneous solution of a given theory can be formally related via a disformal transformation to  a homogeneous solution of another theory~\cite{Bakopoulos:2023sdm, Charmousis:2025xug}.

 So let us concentrate on homogeneous solutions with $Z=-1$. The second  equation above, Eq.~\eqref{easy2}, is an algebraic equation for $X$ and effectively yields the scalar field profile, while \eqref{easy3} gives us the metric component. Using these integrability properties, 
 explicit solutions have been obtained for a family of  DHOST theories \eqref{DHOSTaction} whose associated functions\footnote{Using the Beyond Horndeski notation, we have
    $$G_2(X)=-\frac{2 \alpha}{ \lambda^2} X^\nth,\; \quad
    G_4(X)=1- \alpha X^\nth,\; \quad F_4(X)=\frac{\alpha}{4}(2\nth-1) X^{\nth-2} \, .$$.} take the form
\begin{align}
\label{P_n}
    P(X)&=-\frac{2 \alpha}{ \lambda^2} X^\nth \, , \quad
    \Ftwo(X)=1- 2X A_1(X)\, , \quad
    A_1(X) =\frac{\alpha}{2} X^{\nth-1}\,, \quad 
A_3(X)=\frac{\alpha}{2}(2\nth-1) X^{\nth-2}\,,
\qquad
\end{align}
while $A_2=-A_1$, $A_4=-A_3$ and $A_5=0$, thus  belonging to beyond Horndeski theories according to \eqref{Ai_GLPV}.
These theories are characterised by three parameters: $\nth$   takes positive integer or half-integer values henceforth, $\alpha$  is a dimensionless coupling constant and $\lambda$  is a constant with the dimension of length.

For this family of theories,  the second field equation \eqref{easy2} leads to
the following scalar field profile:
\begin{equation}
X_0=\frac{q^2 /2}{1+(r / \lambda)^2} \qquad\implies \qquad
\psi^{\prime}(r)^2=\frac{q^2}{\cA(r)^2}\left[1-\frac{\cA(r)}{1+(r / \lambda)^2}\right] \,. \label{scalar}
\end{equation}
The last equation \eqref{easy3} is a simple ordinary differential equation for the metric component $\cA$.
It can be solved for the family of theories \eqref{P_n} considered here and, using \eqref{scalar}, one finds
\begin{equation}
    \cA(r)=1-\frac{2\mu}{r}-\xi_\nth\frac{2\lambda}{r} \,\Xi_\nth(r/\lambda)\,,\qquad \Xi_\nth(x)\equiv\int_0^x \mathrm{d}u \; \frac{u^2}{\left(1+u^2\right)^\nth}\,,
    \label{solAq0}
\end{equation}
where $\mu$ is an integration constant (with dimension of length,  since we work implicitly in units where $G=1$ and $c=1$) and  $\xi_\nth$ is a dimensionless parameter, 
\begin{equation}
\label{xi}
    \xi_\nth\equiv \alpha ({2\nth-1}) \left({q^2}/{2}\right)^{\nth} \,,
\end{equation}
which can be interpreted as the ``strength'' of the scalar hair.
The integral $\Xi_\nth$ can be expressed in terms of a hypergeometric function \cite{Baake:2023zsq,Bakopoulos:2023sdm},
\begin{equation}
\Xi_\nth(x)=\frac{x^3}{3}\, {}_2F_1(3/2,\nth;5/2;-x^2)\,,
    \label{hyper}
\end{equation}
which reduces to standard functions for special values of $\nth$, as will be illustrated below. When $x\to +\infty$, $\Xi_p$ behaves as 
\begin{equation}
    \Xi_\nth(x)= \frac{\sqrt{\pi}\,\Gamma(\nth-\frac32)}{4\, \Gamma(\nth)}+{\cal O}\left(\frac{1}{x^3}\right)+x^{-2\nth}\left(\frac{x^3}{3-2\nth}+\frac{\nth}{2\nth-1}x-\frac{\nth(1+\nth)}{2(1+2\nth)x}+{\cal O}\left(\frac{1}{x^3}\right)\right)\,,
\end{equation}
for $\nth\neq -1/2, 1/2, 3/2$ (the special case $\nth=1/2$ is discussed below). In this asymptotic regime, it is  convenient to combine the constant term of $\Xi_\nth(r/\lambda)$  with $\mu$ so as to define the constant
\begin{equation}
\label{ADM}
    M\equiv \mu+\frac{\sqrt{\pi}\,\Gamma(\nth-\frac32)}{4\, \Gamma(\nth)}\xi_\nth\lambda \qquad (\nth\neq -1/2, 1/2, 3/2)\,,
\end{equation}
which can be interpreted as the usual ADM mass of the solution.

In summary, setting
\begin{equation}
    \check{\Xi}_\nth\equiv \frac{\mu -M}{\xi_\nth\lambda}+\Xi_\nth,
\end{equation}
we have  obtained a solution with primary hair, of the form
\begin{equation}
    \cA(r)=1-\frac{2M}{r}-\xi_\nth\frac{2\lambda}{r} \, \check{\Xi}_\nth(r/\lambda)\,,
    \label{solAq}
\end{equation}
parametrised by the mass $M$ and the constant $\xi_\nth$, which quantifies the deviation of the solution from the standard Schwarzschild metric (recovered in the limit $\xi_\nth=0$).
Note that the dimensionless coupling constant $\alpha$  only enters through its sign, since it combines with the primary hair parameter $q$ to  form the dimensionless constant $\xi_{\nth}$ (defined in \eqref{xi}), which can differ for every compact object.
By contrast, the length scale $\lambda$,  fixed by the theory, is universal and sets the characteristic scale at which  the scalar field affects the
geometry, as can be seen from the expression of $X_0$ in \eqref{scalar}, suppressed for $r\gg \lambda$.

If $\xi_\nth>0$ then the horizon of the black hole is more compact than that of a GR black hole  with the same mass. As we increase the value of $\xi_\nth$ (with fixed  mass $M$),  an inner horizon appears, while the event horizon continues to shrink, in a very similar fashion to an electric or magnetic Reissner-Nordstrom BH. For $\nth>3/2$, one can also define a Noether charge associated with the shift symmetry (see \cite{Charmousis:2025xug}). 

It is worth stressing that the solution \eqref{solAq} is not necessarily a black hole. Indeed,  if the function $\cA$ does not vanish, the solution describes a naked singularity, or a soliton in the particular cases where $\cA$ is regular at the origin. Since $\Xi_\nth(x)\simeq x^3/3$ when $x\to 0$, it is clear from \eqref{solAq0} that the solution is regular for $\mu=0$, i.e. when the mass $M$ takes the particular value
\begin{equation}
\label{M_reg}
    M^{\rm reg}_\nth\equiv \frac{\sqrt{\pi}\,\Gamma(\nth-\frac32)}{4\, \Gamma(\nth)}\xi_\nth\lambda \qquad (\nth\neq -1/2, 1/2, 3/2)\,.
\end{equation}
If $\cA$ vanishes for some finite radius, then the solution with the mass $M^{\rm reg}_\nth$ yields a regular black hole, i.e. devoid of singularity behind the horizon.

\medskip

Throughout our study we will focus on the  explicit example given by $\nth=2$, which we now briefly describe. 
For $\nth=2$, the theory  is defined by the functions 
\begin{equation}
\label{n2}
P(X)=-\frac{{2 \alpha}}{\lambda^2}X^2, \quad \Ftwo(X)=1-\alpha X^2\, ,\quad A_1(X)=\frac{\alpha}{2} X\,,\quad
A_3(X)=\frac32\alpha\,,
\end{equation}
yielding the metric component
\begin{equation}
\label{A_n2}
\cA(r)=1-\frac{2 M}{r}+\xi_2\left(\frac{\pi / 2-\arctan \frac{r}{\lambda}}{r/\lambda }+\frac{1}{1+\frac{r^2}{\lambda^2}}\right) \, .
\end{equation}
The solution becomes regular when the mass $M$ takes the special value
\begin{equation}
\label{Mreg2}
    M^{\rm reg}_2=\frac{\pi}{4} \xi_2 \lambda\,,
\end{equation}
which requires $\xi_2>0$. Interestingly, depending on the value of $\xi_2$, the regular solution can be either a black hole, if $\xi_2>  2.8$, or a soliton for smaller values of $\xi_2$ (see \cite{Charmousis:2025xug}).

Note that 
the above metric component \eqref{A_n2} entails a deviation from  standard gravity around a spherical object. A detection of this deviation, for example by measuring the motion of small objects {\it close to} a BH would in principle enable one to measure $\xi_2$ and $\lambda$. The BH mass (deduced from the attraction of {\it distant} objects  by the BH) should then satisfy the consistency inequality $M_{\rm BH} > (\pi/4) \xi_2\lambda$  because a smaller mass would correspond to a naked singularity.\footnote{This is strictly true only for $\xi_2<  2.8$. For higher values of $\xi_2$, the inequality will be slightly different (see the phase diagram in \cite{Charmousis:2025xug}).}

\section{Equations of motion}
\label{section_EoM}

\subsection{Reduced action and equations of motion}
A convenient way to derive the equations of motion is to compute the reduced action obtained by substituting the metric \eqref{background_metric} into the action \eqref{DHOSTaction}, with the choice of DHOST functions \eqref{P_n}. After some integrations by parts, we get the action 
\begin{align*}
    S_{\rm grav}[\cA,\cB,X]=&\frac{2\pi}{\kappa}\int \mathrm{d}t\,\,  \mathrm{d}r\, \sqrt{\frac{\cA}{\cB}}\Biggl\{2\left(1-\cB-r \cB'\right)\nonumber\\
    &  +\alpha X^p \left[ - 2 \left( 1 + \frac{r^2}{\lambda^2} \right) + 4 \cB - \frac{q^2}{X} \frac{\cB}{\cA}+2r \left(\cB \frac{\cA'}{\cA} +\cB' +2\nth \frac{X'}{X} \frac{\cB}{\cA} \left(\cA+ \frac{1-2\nth}{4\nth}\, \frac{q^2}{X} \right)\right)\right]\Biggr\}.
\end{align*}
The above gravitational action must be supplemented by the action $S_{\rm m}$ for the matter, which is  assumed to be minimally coupled to the metric. The variation of $S_{\rm m}$ with respect to the metric defines the matter energy-momentum tensor $T^{\mu \nu}$. Here, we model the matter as a barotropic perfect fluid, which is a very good approximation for a neutron star, so that 
\begin{equation}
    T^{\mu \nu} \equiv  \frac{2}{\sqrt{-g}} \frac{\delta S_{\rm m}}{\delta g_{\mu \nu}}=(\rho + \p)u^{\mu}\,u^{\nu}+\p g^{\mu \nu}\,,
\end{equation}
where $\rho$ is the energy density and $\p$ the pressure.
The conservation of the energy-momentum tensor, 
\begin{equation}
    \nabla_\mu T^{\mu}_{\ \nu}=0\,,
\end{equation}
yields, for the metric \eqref{background_metric}, the relation 
\begin{equation}
    \p'= -\frac{\cA '}{2\cA}(\rho+\p)\,,
    \label{eP}
\end{equation}
which is convenient to use instead of the angular components of the metric equations.

Varying  the total action $S=S_{\rm grav}+S_{\rm m}$ with respect to $\cA$ and $\cB$,  we get 
\begin{subequations}
    \begin{align}
         \cA'&= \frac{\cA}{\cB}\left[\frac{\left(1-\cB\right)}{r}+\kappa r(\p-\rho_{\rm eff})\right],\label{eA}
    \\
    \cB'&= \frac{1-\cB}{r}-\kappa \,r (\rho+\rho_{\rm eff}),\label{eB}
    \end{align}\label{system A B}
\end{subequations}
where we have regrouped all the beyond-GR terms in an effective energy density, defined by
\begin{equation}\label{rhoeff0}
   \kappa \rho_{\rm eff}\equiv     \xi_\nth \X^{\nth-1} \frac{\cB}{\cA \,r} \left[\frac{\cA  \frac{\X}{\X_0}-\cB}{ (2 \nth-1) r \cB }-\frac{\X'}{\X}  \right]\,,
\end{equation}
using the convenient normalized kinetic density $\X$, and its exterior value $\X_0$ (obtained from \eqref{scalar}): 
\begin{equation}
    X \equiv \frac{q^{2}}{2}\X\,\qquad \implies\qquad\X_0=\frac{1}{1+\frac{r^2}{\lambda^2}}\,.
\end{equation}
The system \eqref{system A B} together with \eqref{eP} has exactly the same form as in GR, except that the usual matter density $\rho$ is now replaced by $\rho+\rho_{\rm eff}$, whereas the pressure is replaced by $P-\rho_{\rm eff}$. This implies that the deviations from  standard GR  can be described as an effective  perfect fluid  energy-momentum tensor with  equation of state 
\begin{equation}
    P_{\rm eff}=-\rho_{\rm eff}\,,
\end{equation}
where both terms are radially dependent.

Finally, varying the action $S$ with respect to $X$ and eliminating the metric derivatives $\cA'$ and $\cB'$ using \eqref{eA} and \eqref{eB}, we obtain
\begin{equation}
  \X = \frac{\cB}{\cA}\,\Theta \,\X_0,\qquad \Theta\equiv 1-\frac{2\nth-1}{2\nth\,\cB}\, r^2\,\kappa(\rho+\p)
  \,,\label{Xbar}
\end{equation}
which effectively extends the expression \eqref{scalar} to the interior of the star. Note that when $\nth>1/2$ the effect of the star matter can render $\Theta<0$ and therefore $\X<0$, in stark contrast with BH solutions where $\X\equiv \X_0>0$.  

Using \eqref{Xbar}, we can rewrite \eqref{rhoeff0} in the form 
\begin{equation}
\label{rho_eff_bis}
    \kappa \rho_{\rm eff}\equiv     -\xi_\nth \X^{\nth-1} \frac{\cB}{\cA \,r} \left[\frac{\kappa r(\rho+P)}{2p \cB}+\frac{\X'}{\X}  \right]\,.
\end{equation}
The system \eqref{eP}-\eqref{system A B} appears deceivingly simple and hides the complexity arising from the scalar field's influence  on the stellar profile. Indeed, according to \eqref{rho_eff_bis} and \eqref{Xbar}, the quantity $\rho_{\rm eff}$ contains derivatives of $\cA$ and $\cB$, which must be transferred  to the left-hand side of the equations to properly formulate a first-order system. In the next subsection, we will explicitly disentangle  these terms.

\subsection{Impact of the scalar field}

Before analysing in detail the new terms in the equations of motion, let us first substitute, as is traditional in GR (see for example \cite{schutz2022first}), the local mass function $m(r)$ for the metric component $\cB$, so that 
\begin{equation}
\label{mass_local}
    \cB=1-\frac{2m(r)}{r}\,.
\end{equation}
The equations \eqref{eA} and \eqref{eB} then become
\begin{equation}
\label{m'}
m'=\frac{r^2}{2}\kappa (\rho+\rho_{\rm eff})    
\end{equation}
and 
\begin{equation}
    \frac{\cA'}{\cA}=\frac{2m+ r^3 \kappa(P-\rho_{\rm eff})}{r(r-2 m)}\,.
\end{equation}
Substituting this latter equation into \eqref{eP} and using \eqref{m'} to replace $\rho_{\rm eff}$ by $m'$,   we obtain the modified TOV equation{\footnote{In GR, the TOV equation reads
\begin{equation}
 P'=-\frac{\rho+P}{r(r-2m)}\left(m+\frac{\kappa}{2}r^3 P\right)\,.  
\end{equation}}}:
\begin{equation}
 P'=-\frac{\rho+P}{r(r-2m)}\left(m+\frac{\kappa}{2}r^3(P+\rho)- m' r\right) \,.  
\end{equation}

It now remains to evaluate $\X'$ in the star interior from \eqref{Xbar} in order to decrypt $\rho_{\rm eff}$ in full generality. Defining
the quantity
\begin{equation}
    \bb  \equiv
     \xi_\nth\frac{(2 \nth-1)}{4 \nth}\frac{r^2\kappa (\rho +\p)}{\cA^2 \left(1+r^2/\lambda ^2\right)}\X^{\nth-2}\,\label{b0}\,,
\end{equation}
we can rewrite the term proportional to the effective energy density on the right hand side of \eqref{m'} in the form
\begin{equation}
\label{rhoeff}
   \frac{r^2}{2}\kappa  \rho_{\rm eff}=\frac{\xi_\nth \bar{X}^p}{\lambda^2 \Theta}r^2-(3+\frac{1}{c_s^2})\bb\left[\frac{m}{r}-m'+\frac{ r^2 \kappa}{2}(\rho+P)\right]+\bb\left[3(1-\frac{2m}{r})+\frac{ r^2\kappa}{2\nth}(\rho+P)\right]\,,
\end{equation}
where we have introduced the sound speed $c_{\rm s}^{2}\equiv {\p'}/{\rho'}$.
Having obtained $\rho_{\rm eff}$ in an expanded form, it is now straightforward to write down the modified field equations  explicitly as a first order ODE system of three equations governing the local mass $m$, the metric function $\cA$ and the pressure $\p$:
\begin{subequations}
\begin{align}
    m'&=\frac{1}{1-(3+\frac{1}{c_s^2})\bb}\left[\frac{\kappa r^2 \rho}{2} + \xi_{\nth} \frac{ \bar{X}^{p}}{\Theta \lambda^2}r^2-(3+\frac{1}{c_s^2})\frac{\bb}{r}\left(m+\frac{\kappa r^3}{2}(\rho+P)\right)\right.\nonumber\\
   &\left.\qquad\qquad \qquad\qquad\qquad\qquad\qquad\qquad\qquad\qquad\qquad
   +\frac{\bb}{r}\left(3(r-2m)+\frac{\kappa r^3(\rho+P)}{2p}\right)
    \right]
    ,\label{eBNEW1}\\
    \frac{\cA'}{2\cA}&= \frac{1}{(r^2-2mr)[1-(3+\frac{1}{c_s^2})\bb]}\left[m+\frac{\kappa r^3 P}{2}-\xi_{\nth} \frac{ \bar{X}^{p}}{\Theta \lambda^2}r^3
    -\bb\left(3(r-2m)+\frac{\kappa r^3(\rho+P)}{2p}\right)\right],\label{eANEW2}
    \\
     \p'&= -\frac{(\rho+P)}{(r^2-2mr)[1-(3+\frac{1}{c_s^2})\bb]}\left[m+\frac{\kappa r^3 P}{2}-\xi_{\nth} \frac{ \bar{X}^{p}}{\Theta \lambda^2}r^3
    -\bb\left(3(r-2m)+\frac{\kappa r^3(\rho+P)}{2p}\right)\right]\,,\label{ePNEW3}
    \end{align}
\end{subequations}
where $\bb$ is given by \eqref{b0}, and 
\begin{equation}
\label{Theta}
    \Theta=1-\frac{(2\nth-1)r^2}{2\nth\,(1-\frac{2m}{r})}\kappa(\rho+\p)\,,\qquad \bar{X}=\frac{2\nth(1-\frac{2m}{r})-(2\nth-1)r^2\kappa(\rho+\p)}{2\nth\,\cA(1+\frac{r^2}{\lambda^2})}\,.
\end{equation}
Note that the above equation for $P'$ directly follows from the substitution of \eqref{eANEW2} into the energy conservation equation \eqref{eP}. This is the final system of equations, which we will solve numerically in  Section \ref{section_numerics}.

Before discussing, in the next subsection, a few specific features of the above system,  it is worth noting that in the particular case $p=1/2$, which corresponds to a Horndeski theory, one recovers exactly  the same TOV equations as in GR, despite the presence of a non trivial scalar field. Indeed, in this case, the  effective energy density $\rho_{\rm{eff}}$ vanishes since $\xi_\nth$, defined in  \eqref{xi}, and  $\tau$, defined in \eqref{b0}, are zero. Furthermore, the star's exterior is described by a stealth Schwarzschild metric \cite{Bakopoulos:2023fmv} --- the unique Horndeski homogeneous spacetime where the primary hair dependence \eqref{xi} drops out, giving a pure GR geometry with a non trivial scalar \eqref{scalar}. In summary the ``stealth'' property for $p=1/2$ extends from the vacuum solution  to the neutron star solution. Since, at the background level at least, the star  is identical to its GR counterpart,  we will not consider this case further in our discussions.

\subsection{Characteristic properties and singularity for the neutron star interior}
\label{subsection_singularity}
The star's profile obeys the first order system \eqref{eBNEW1}-\eqref{ePNEW3} which has some special  properties as compared to the equivalent system in GR.

Firstly, we note that in the absence of matter ($\rho=P=0$) all the $\bb$ dependent terms in  the expression \eqref{rhoeff} for $\rho_{\rm eff}$ drop out,  leaving nevertheless the first term which continues to contribute to the equations of motion and in particular to the ADM mass for the vacuum  solution \eqref{ADM}. In other words, the integrated mass $m(R)$ at the surface of the neutron star, where the radius $R$ is defined by $P(R)=0$, can be  different from the ADM mass as measured by an observer at spatial infinity for this class of theories \eqref{P_n}. This effect, which is of course absent in GR, depends on the sign and magnitude of the scalar charge $\xi_\nth$ and on the range of $\lambda$.

Secondly, we observe that  the system \eqref{eBNEW1}-\eqref{ePNEW3} can become become singular if 
\begin{equation}
   \tilde\tau\equiv (3+\frac{1}{c_s^2})\bb=1\,,
   \label{b01}\,,
\end{equation}
in which case the common denominator on the right-hand side of all three equations vanishes. 
From \eqref{b0}, this is only possible for  $\xi_\nth>0$, since $\rho+\p\geq 0$.
For ordinary stars, like the sun, $\bb$ is expected to be very small, but for a very compact star, with the metric component $\cA$ becoming small, the function $\tilde\tau$ can approach, or even reach, unity.
Even if $\tau$ remains small, the singularity can also occur in a region where the sound speed becomes very low.

A third important point concerns the identical square brackets of \eqref{eANEW2} and \eqref{ePNEW3} which, unlike the 
condition \eqref{b01}, 
are independent of the sound speed $c_s^2$.
Since $\rho+\p\geq 0$, in GR the pressure $\p$ is always  a monotonically decreasing function from the star's center to its surface. For $\xi_\nth>0$, however, this behaviour can change dramatically:  the pressure may first increase, reaching a local maximum in the star's interior (with $\cA$ simultaneously attaining a local minimum at the same point) before eventually decreasing.

For a generic $p$, such a behavior can be deduced from the radial expansion  of the pressure near the star's center (see Appendix \ref{appB}):
\begin{equation}
    \p=\p_c -\left(\p_c+\rho _c\right)\left[\frac{\kappa }{12}   \left(3 \p_c+\rho _c\right)
    -\frac{ \xi _\nth }{\cA_c^{\nth}} \left(\frac{2 \nth-1}{4\nth} \kappa  (\rho _c+  \p_c)+\frac{1}{3\lambda ^2 }\right)\right]r^2+\dots \,.
\label{P_expansion}
\end{equation}
Whereas the coefficient of $r^2$ is always negative in GR, it can become positive if the extra term, proportional to $\xi _\nth>0$, dominates. The effect becomes more pronounced for larger $\xi_\nth$, higher central density $\rho_c$   (and thus smaller $\cA_c$) or smaller $\lambda$. Such non-monotonic  profiles for the matter pressure and energy density might lead to instabilities under radial perturbations.

Let us  go one step further and estimate the influence of the parameters $\xi_2$ and $\lambda$ within the squared brackets in the case $p=2$. Using our definitions  \eqref{Xbar} and \eqref{b0} the modified TOV equation \eqref{ePNEW3} can be written as
\begin{equation}
   \p'= -\frac{(\rho+P)}{(r^2-2mr)(1-{\tilde \tau})}\left[m+\frac{\kappa r^3 P}{2}\right]+\Q \,,
\end{equation}
where the quantity
\begin{eqnarray}
\label{Q}
    \Q\equiv\frac{(\rho+\p)\X_0 }{3r(1-{\tilde \tau}) }\left[\frac{r^2}{\lambda^2}\frac{\Theta^2-5\Theta+10}{1-\Theta}+10-\Theta\right] \bb
\end{eqnarray}
englobes 
all the beyond GR effects, except the denominator $(1-{\tilde \tau})$.
Assuming here that ${\tilde \tau}<1$ and  $\xi_2>0$,  we observe that the additional term $\Q$
competes with the GR term and can disrupt  the monotony of $\p$. 
Since $\Theta\leq 1$, the term in square brackets in \eqref{Q} is positive and hence small $\lambda$ or positive large $\xi_2$ always enhance the magnitude of $\Q$,  further modifying the  hydrostatic equilibrium 
relative to GR.

\begin{figure}[h] 
\centering
\includegraphics[scale=0.5]{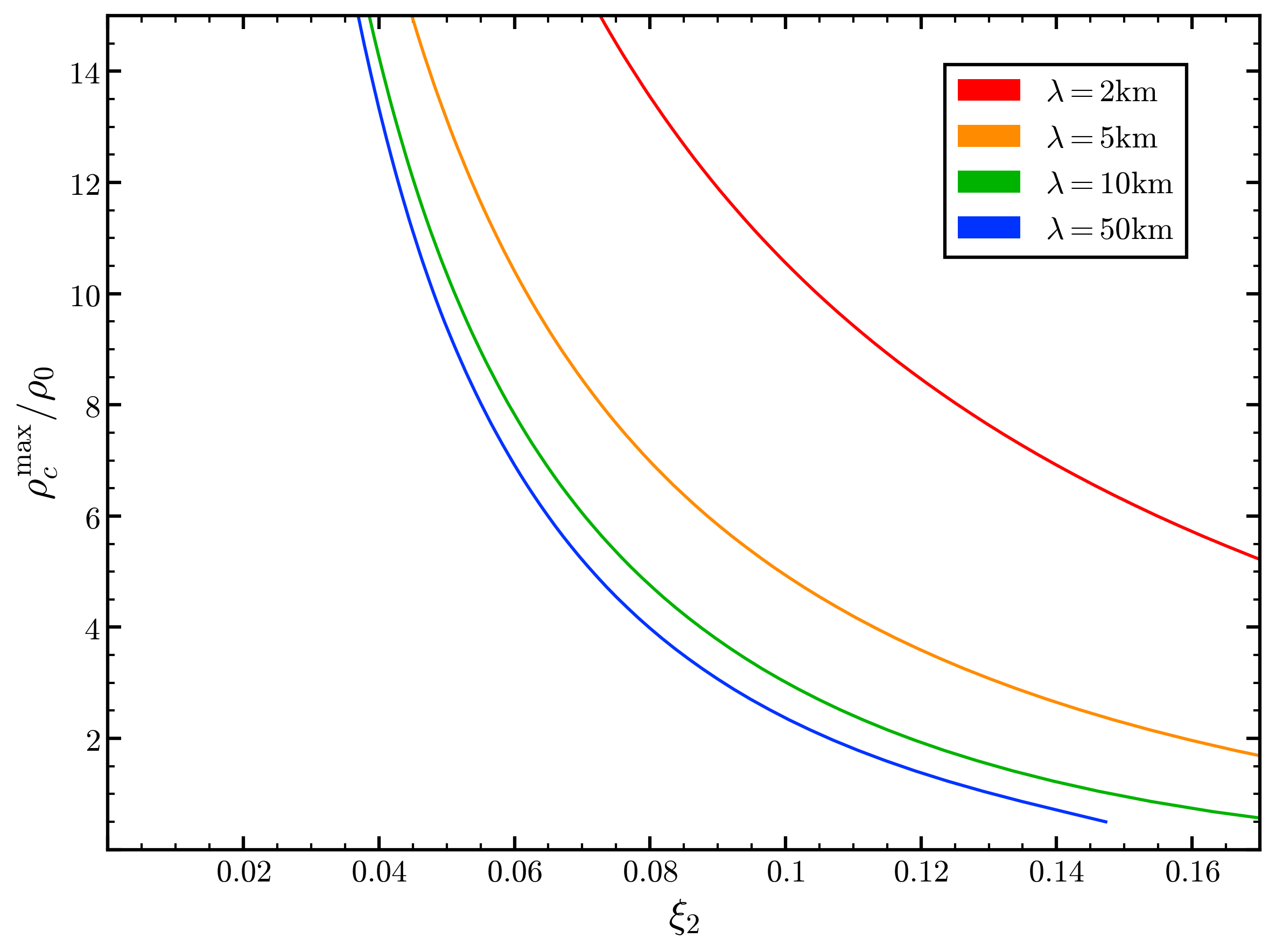}
\caption{Upper bounds in the parameter space $(\xi_2,\rho)$, defined by the maximal central density $\rho_c^{\rm max}$ allowed for each $\xi_2$, with $\lambda=2,5, 10$ and $50$ km. The plot corresponds to the polytropic EoS defined in \eqref{EOSpolytrope}-\eqref{polyc}.}
\label{pxirhomax_poly}
\end{figure}

Given the above discussion we expect to find upper bounds in the parameter space spanned by the parameters $\xi_2$ and $\rho_c$, as confirmed by our numerical calculations. The precise limits depend on the value of $\lambda$ and on the equation of state. In Fig.~\ref{pxirhomax_poly}, we have plotted these boundaries, assuming a polytropic EoS (defined explicitly in the next section), for four different values of the length parameter $\lambda$. Note that for higher values of $\lambda$, the curves tend to coalesce.
\medskip

\section{Numerical analysis}
\label{section_numerics}
To understand in more detail how the internal structure of  compact stars is affected by the gravitational theories considered here, we integrate numerically equations  (\ref{eBNEW1}), (\ref{eANEW2}) and (\ref{ePNEW3}), supplemented by an EoS. 
We start with the simple case of a polytropic EoS before progressing to more complex and realistic EoS later on.

\subsection{Neutron star structure for a polytropic equation of state}

We first consider,  for simplicity, a polytropic EoS, which can be written in the following dimensionless form, expressing the pressure and energy density in terms of the particle number density $n$:
\begin{equation}
    \frac{\p}{\rho_0 c^2}= K\, \left(\frac{m_b\, n}{\rho_0}\right)^\Gamma,\quad
    \frac{\rho}{\rho_0}=\frac{m_b \,n}{\rho_0}+ \frac{K}{\Gamma-1}\left(\frac{m_b\,  n }{\rho_0}\right)^{\Gamma}\,,\label{EOSpolytrope}
\end{equation}
where $m_b=1.6749 \times 10^{-24}$g  is the mean baryonic mass and   $\rho_0= m_b n_0 $ ($n_0=0.1\, {\rm fm^{-3}}$ being the typical nuclear number density) is our reference  density. In our numerical computations, we have taken the following values for the constant  $K$ and the adiabatic index $\Gamma$ :
\begin{equation}
\label{polyc}
K=0.043 \,, \qquad  \Gamma =5/3\,.  
\end{equation}

The gravitational theory depends on  the value of $p$, which we choose to be $p=2$, and the length scale parameter $\lambda$, which corresponds to the typical scale on which the scalar interaction is efficient. 
Finally, the amplitude 
of the beyond-GR effects is quantified by the parameter $\xi_2$, which in principle can vary from one object to another, in contrast to $\lambda$.
Therefore, once the EoS and the theory are fixed, the neutron star depends on only two parameters: the first one, as in GR, is the central density $\rho_c$ (or $n_c$), while  the  second one, $\xi_2$,  represents in some sense the ``charge'' of the star.

Given the above  EoS and a specific choice of  parameters, we compute the radial  profiles of the matter and of the metric by  integrating Eqs (\ref{eBNEW1}), (\ref{eANEW2}) and (\ref{ePNEW3}). 
The integration is performed from  the  center of the star, where we  impose usual regularity conditions, as discussed in Appendix \ref{appB}. The only input is  the central energy density $\rho_c$ or, equivalently, the central number density $n_c$, the central pressure being inferred from the equation of state (\ref{EOSpolytrope}). The  system is  numerically integrated  up to the radial point $r = R$ where the pressure vanishes,
\begin{equation}
\p(R) = 0,
\end{equation}
thus defining $R$ as the radius of the star. Note that our choice $\Gamma =5/3$ for the polytropic index prevents the appearance of  discontinuities at the surface of the star\footnote{Near the surface of the star, we have $\tilde\bb\sim \rho^{2-\Gamma}/(\Gamma K)$  for $\Gamma>1$. If $\Gamma<2$, then  $\tilde\bb\to 0$, avoiding  a discontinuity of the metric and scalar field at the star's surface (see e.g. \cite{Kobayashi:2025bdh,Kobayashi:2018xvr}).}. 
The central value of the metric component $\cA$, denoted $\cA_c$, is determined  by imposing the boundary condition at spatial infinity $\cA(\infty)=1$. 

\begin{figure}[H] 
\centering
\includegraphics[scale=0.5]{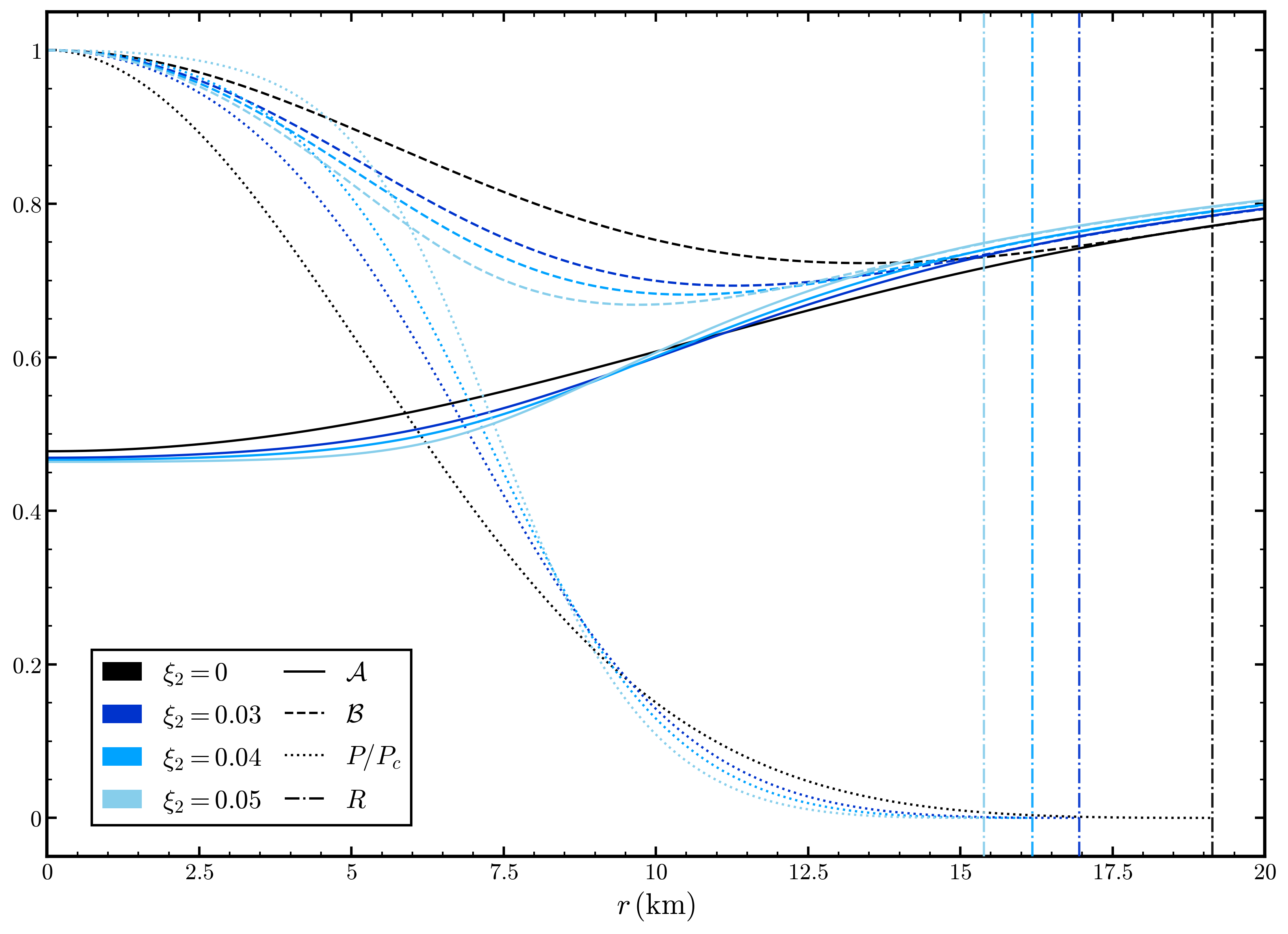}
\caption{Radial profiles of  $\cA$, $\cB$ and $\p$ for the polytropic EoS given by \eqref{EOSpolytrope} and \eqref{polyc}, assuming $n_c= 4\, n_0$ and  $\lambda=50\,{\rm km}$. Here we focus on the effect of $\xi_2$, plotting the profiles for $\xi_2=0$ (GR theory, in black), $\xi_2=0.03,0.04$ and $0.05$ (in shades of blue).}
\label{pABrhoP}
\end{figure}

 Fig.~\ref{pABrhoP} illustrates how the internal structure of neutron stars is affected by increasing the value of $\xi_2$, starting from  $\xi_2=0$ which corresponds  to  GR.  As $\xi_2$ increases, one observes that the pressure and energy density in the star's core  decrease less rapidly than in GR, before decreasing much more rapidly in the outer layers of the star.  The star radius, indicated by a vertical  line,  is smaller for higher values of $\xi_2$. For the metric component $\mathcal{A}$, the increase is slower in the star's core   than in GR, but accelerates outward, ultimately exceeding its GR value.  The  component $\mathcal{B}$ first  decreases more sharply than in GR, then rises more rapidly, before joining $\mathcal{A}$ at the surface.  We have checked that the numerical integration outside the star reproduces the analytical solution \eqref{A_n2}.

\begin{figure}[h] 
    \centering
    \includegraphics[scale=0.45]{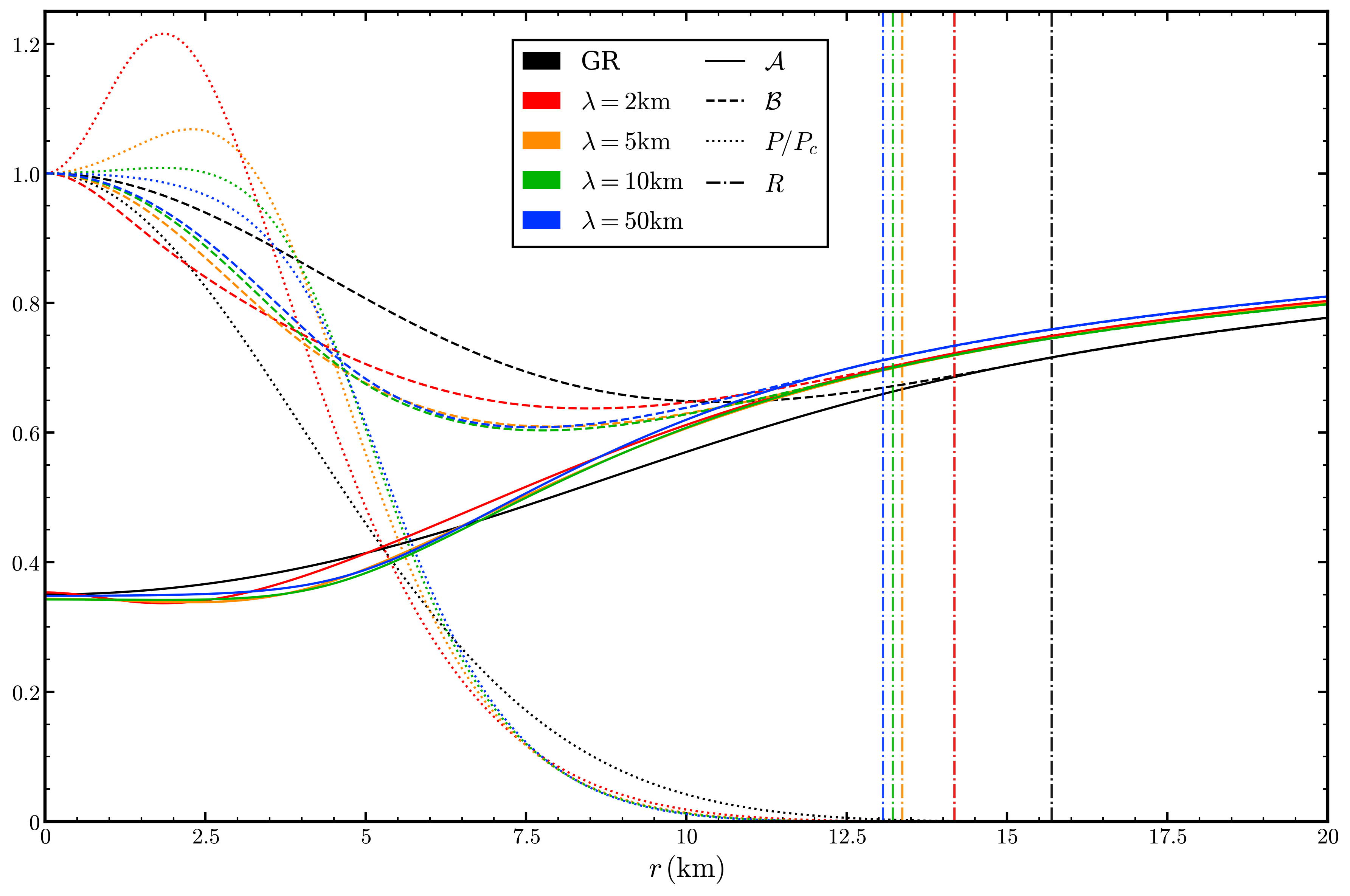}
    \caption{Radial profiles of  $\cA$, $\cB$ and $\p$ for the polytropic EoS, assuming $n_c= 8\, n_0$ and  $\xi_2= 0.03$. Here we focus on the effect of $\lambda$, plotting the profiles for $\lambda=2,5, 10$ and $50$ km. These are to be compared with the GR case pictured in black.}
    \label{pABrhoP22}
\end{figure}

As a further exploration, we now study how the length parameter $\lambda$ affects the neutron star profile. We have plotted in Fig.~\ref{pABrhoP22} several configurations with the same scalar charge $\xi_2$ and central density $\rho_c$, varying $\lambda$. A notable
feature is that, for sufficiently small $\lambda$, stars can exhibit a pressure (and thus matter density) profile that initially increases  outward from the center,  reaches a maximum, and then decreases in the outer layers. This is in agreement with our comments on the equations of motion made at the end of Section \ref{section_EoM}.

\begin{figure}[h!]
    \centering
\includegraphics[scale=0.6]{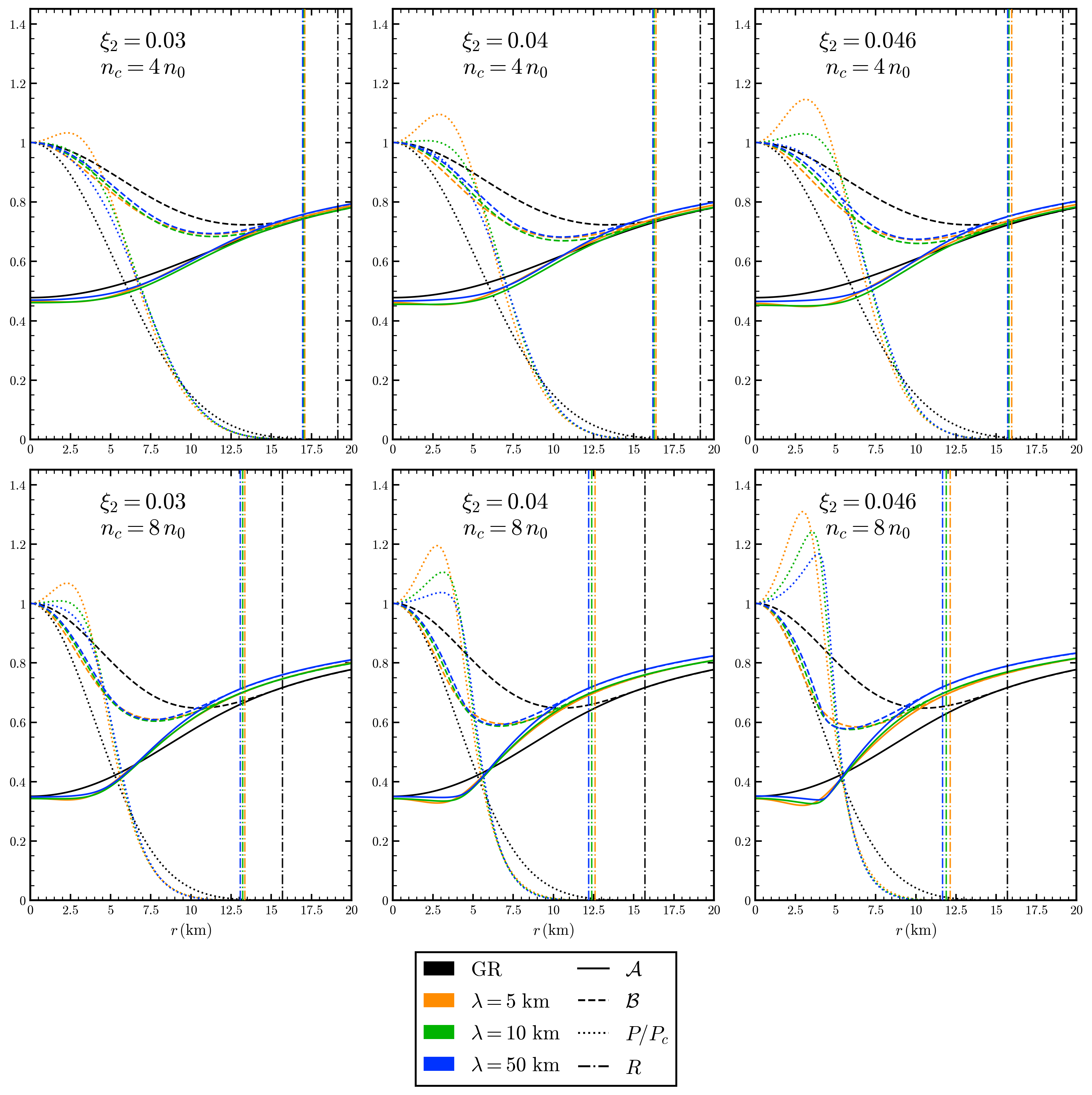}
    \caption{Radial profiles of  $\cA$, $\cB$ and $\p$  for the central densities  $n_c= 4\, n_0$ (top) and $n_c= 8\, n_0$ (bottom), using the polytropic EoS. Here, we study the combined effect of $\lambda$ and $\xi_2$,  for $\lambda=5, 10$ and $50$ km; from left to right, $\xi_2$ is taken to be $0.03$, $0.04$ and $0.046$ (close to the critical value $0.047$  where the singularity appears). 
    }
    \label{fig_profiles}
\end{figure}

As shown in Fig.~\ref{fig_profiles} --- and consistent with Eq.~\eqref{P_expansion} --- the specific features discussed above are more  pronounced for larger $\xi_\nth$, higher central density (i.e. more relativistic stars), or smaller $\lambda$. 
Indeed, as shown in the top panels, it is easier to find a ''pressure peak'', instead of a monotonic pressure,  for smaller  $\lambda$ or larger $\xi_\nth$. The radius corresponding to the maximum of the pressure also tends to be closer to the center of the star when the amplitude of the peak increases, as illustrated by the last figure at the bottom.
One also notes that the metric component $\cA$ decreases in the central region, again in contrast with the usual GR behaviour, but in accordance with the standard conservation equation \eqref{eP} since $P$ increases.

\begin{figure}[h!]
    \centering
    \includegraphics[scale=0.51]{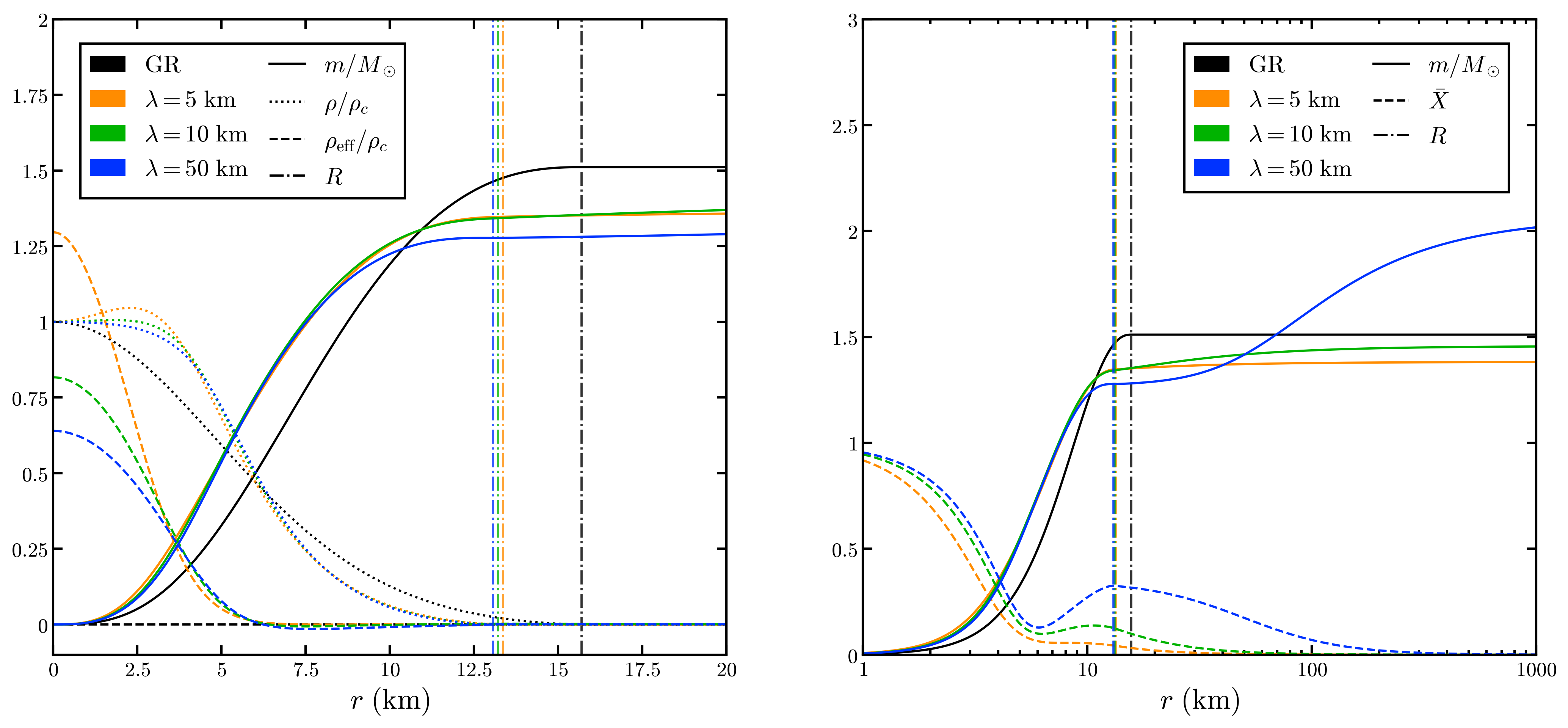}
    \caption{Radial profiles of $m$, $\rho$, $\rho_{\rm eff}$ and $\X$ for $\lambda=5, 10$ and $50$ km, with the polytropic EoS,  $n_c = 8\, n_0$ and $\xi_2=0.03$. The left panel shows the profile of $m$, $\rho$ and $\rho_{\rm eff}$ within the  star, while the right one focuses on the behavior of $m$ and $\X$ at larger distances. Note that for $\lambda=50$ km (blue curve), $m$  continues to increase beyond the star's surface until it reaches its  ADM value far away.}
    \label{prmass}
\end{figure}

It is also instructive to examine the radial profile of the effective energy density $
\rho_{\rm eff}$ inside the star. In  Fig.~\ref{prmass}, we plot  $
\rho_{\rm eff}$,  $\rho$ and the local mass $m$ for  three configurations corresponding to those of Fig.~\ref{pABrhoP22}. For small  $\lambda$, $\rho_{\rm eff}$ increases  significantly  in the star's core, which can be understood from Eq.~\eqref{rho_eff_bis_per}. In the outer layers, however, this trend reverses, with  $\rho_{\rm eff}$  even becoming  slightly negative near the surface. This sign change is explained by Eq.~\eqref{rho_eff_bis}: the term $\X'/\X$ is negative in the core (see the right panel  of Fig.~\ref{prmass}, where $\X$ is plotted) and dominates the first term, yielding  $\rho_{\rm eff}>0$. This is no longer the case in the outer layers,   which explains why $\rho_{\rm eff}$ becomes negative. 

We also observe that, for small values of $\lambda$, typically well below the star radius, the local mass $m$ has the same typical radial evolution as in GR. Indeed, for $\lambda=5$ km in the figure, the mass increases within the star  and  settles to its ADM mass value at the surface. For higher values of $\lambda$  ($\lambda=$ 10 or 50 km in the figure),  however, although the mass $m$ is nearly constant near the surface, it  starts to grow again outside the star, only approaching its ADM value $M$   at larger distances. This difference in behavior  arises from the influence of the scalar field in the exterior region, whose range  increases with $\lambda$  (see Eq.\eqref{Theta}). Moreover, for larger values of $\lambda$,  $M$  becomes larger than in GR. At intermediate radii in the exterior region, the effective mass $m(r)$  depends on the position of the observer.

\newpage
\subsection{Phenomenological equations of state}
\begin{figure}[h]
    \centering
    \includegraphics[scale=0.5]{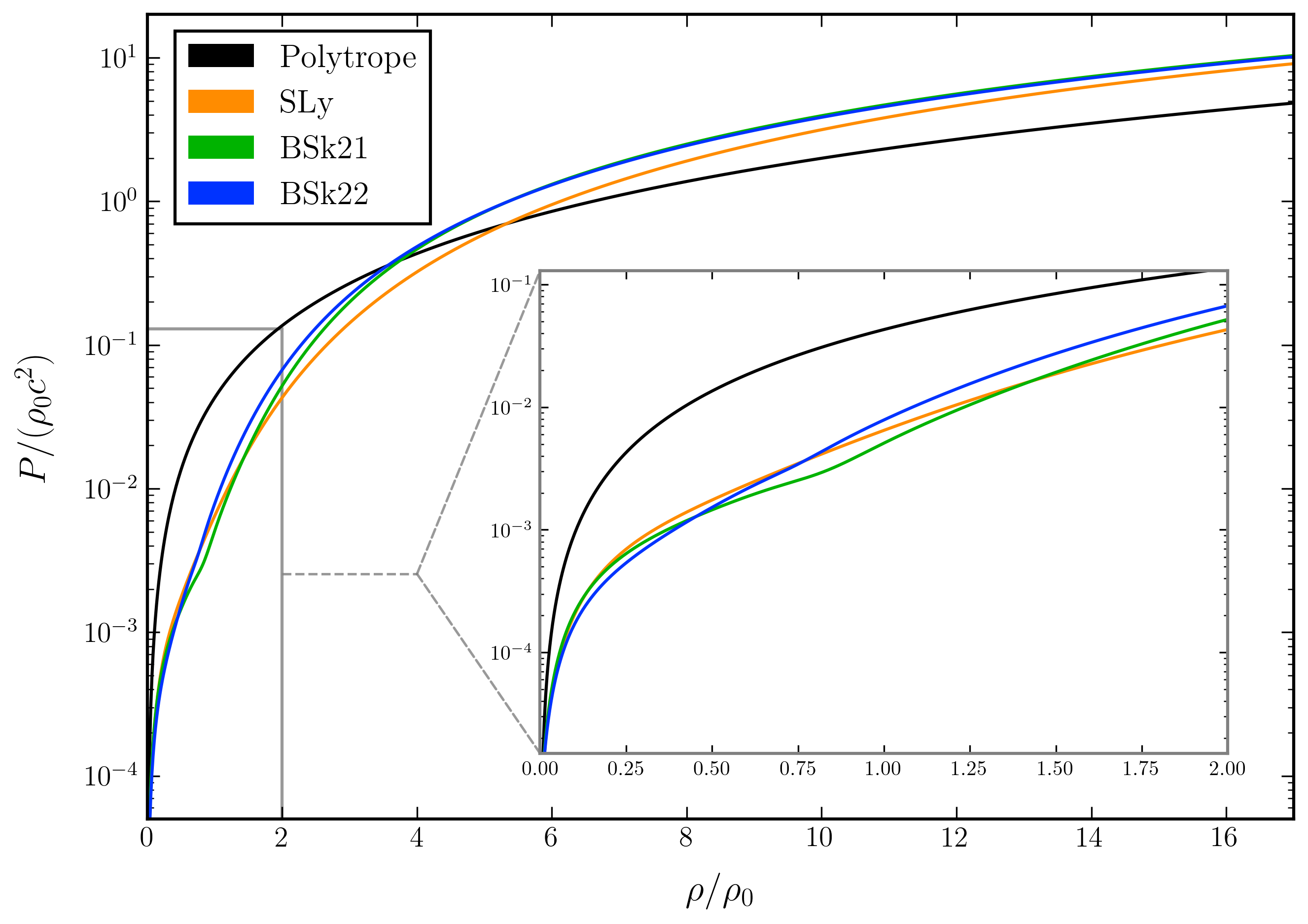}
     \caption{Polytropic, SLy, BSk21 and BSk21 EoS, plotted for $\rho \leq 17 \rho_0$. The  frame zooms in on the curves in the range $[0,2 \rho_0]$.}
    \label{pEOS}
\end{figure}

 We now  consider three more complex EoS commonly used in  neutron star studies: SLy,  BSk21 and BSk22. As  discussed in  \cite{Haensel:2004nu,Potekhin:2013qqa}, these EoS can  be conveniently parametrized in the form 
\begin{eqnarray}
\log_{10}\left(\frac{P/c^2}{{\rm g\,.\,cm}^{-3}}\right) & &= \frac{\tb_1+\tb_2 \chi +\tb_3 \chi ^3}{1+\tb_4 \chi}\;U\bm{[}\tb_5 (\chi -\tb_6)\bm{]}\nonumber\\
& & +(\tb_7+\tb_8 \chi)\;U\bm{[}\tb_9 (\tb_{10}-\chi )\bm{]}+(\tb_{11}+\tb_{12} \chi)\;U\bm{[}\tb_{13} (\tb_{14}-\chi )\bm{]}
\label{EOS}\\ 
& & +(\tb_{15}+\tb_{16} \chi )\; U\bm{[}\tb_{17} (\tb_{18}-\chi )\bm{]}
 +\frac{\tb_{19}}{\tb^2_{20} (\tb_{21}-\chi )^2+1}+\frac{\tb_{22}}{\tb^2_{23} (\tb_{24}-\chi )^2+1}
 \nonumber
\end{eqnarray} with 
\begin{eqnarray}
U\bm{[}x\bm{]}\equiv\frac{1}{e^x+1}\,,\qquad  \chi=\log_{10}({\rho}/(\rm{g}\,.\,{\rm cm}^{-3}))\,.
\end{eqnarray}
Each EoS corresponds to different values of the coefficients $\tb_i$. 
The 18 coefficients for the SLy equation of state\footnote{The coefficients $b_i$  for $19\leq i\leq 24$ vanish in this case.} are given  in  the left table  of App.~\ref{App:tables}, 
 based on \cite{Haensel:2004nu}, while the 24 coefficients for BSk21 and BSk22 are listed 
in the right table of
 App.~\ref{App:tables},
based on
\cite{Potekhin:2013qqa,Pearson:2018tkr}. All three EoS are represented in Fig. \ref{pEOS}, along with the polytropic EoS \eqref{EOSpolytrope}.

\begin{figure}[h!]
    \centering
    \includegraphics[scale=0.45]{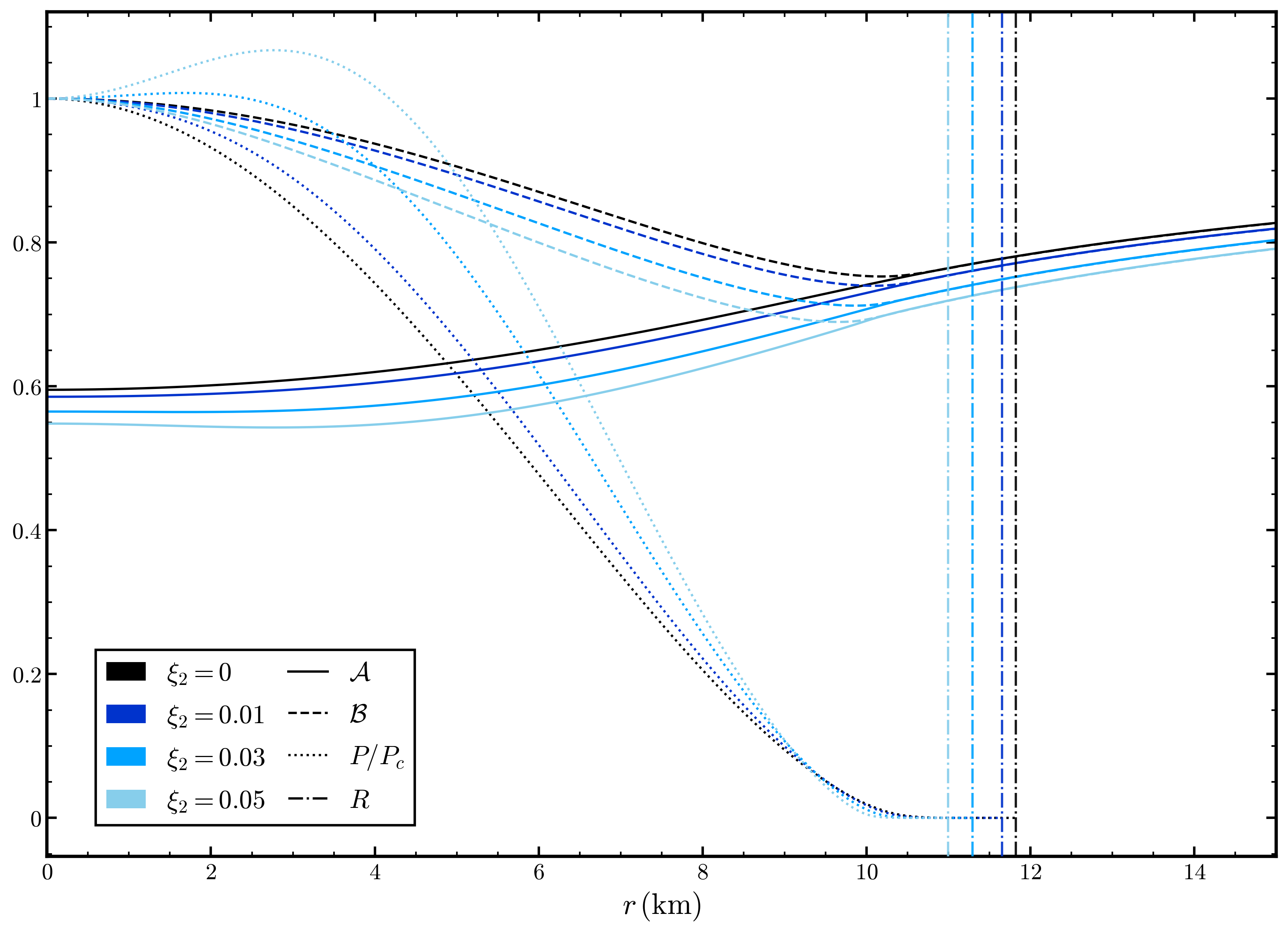}
    \caption{Radial profiles of  $\cA$, $\cB$ and $\p$ for  $\xi_2=0.01,0.03$ and $0.05$ (in shades of blue), with the SLy EoS,  $\rho_c= 4\, \rho_0$ and  $\lambda=5\,{\rm km}$. As before, the GR profile is in black.}
    \label{pSLyprofile}
\end{figure}
In Fig.~\ref{pSLyprofile}
we have plotted the profile of a neutron star with the SLy equation of state for several values of $\xi_2$, fixing $\lambda$ to $5$km and the central density to $\rho_c=4\rho_0$. This figure should be compared with some of the configurations plotted in Fig.~\ref{fig_profiles}. One notes that the pressure profile is monotonic for the lowest (non GR) value of $\xi_2$ but exhibits a maximum at an intermediate radius for the highest values of $\xi_2$.

\begin{figure}[H]
    \centering  \includegraphics[width=0.6\linewidth]{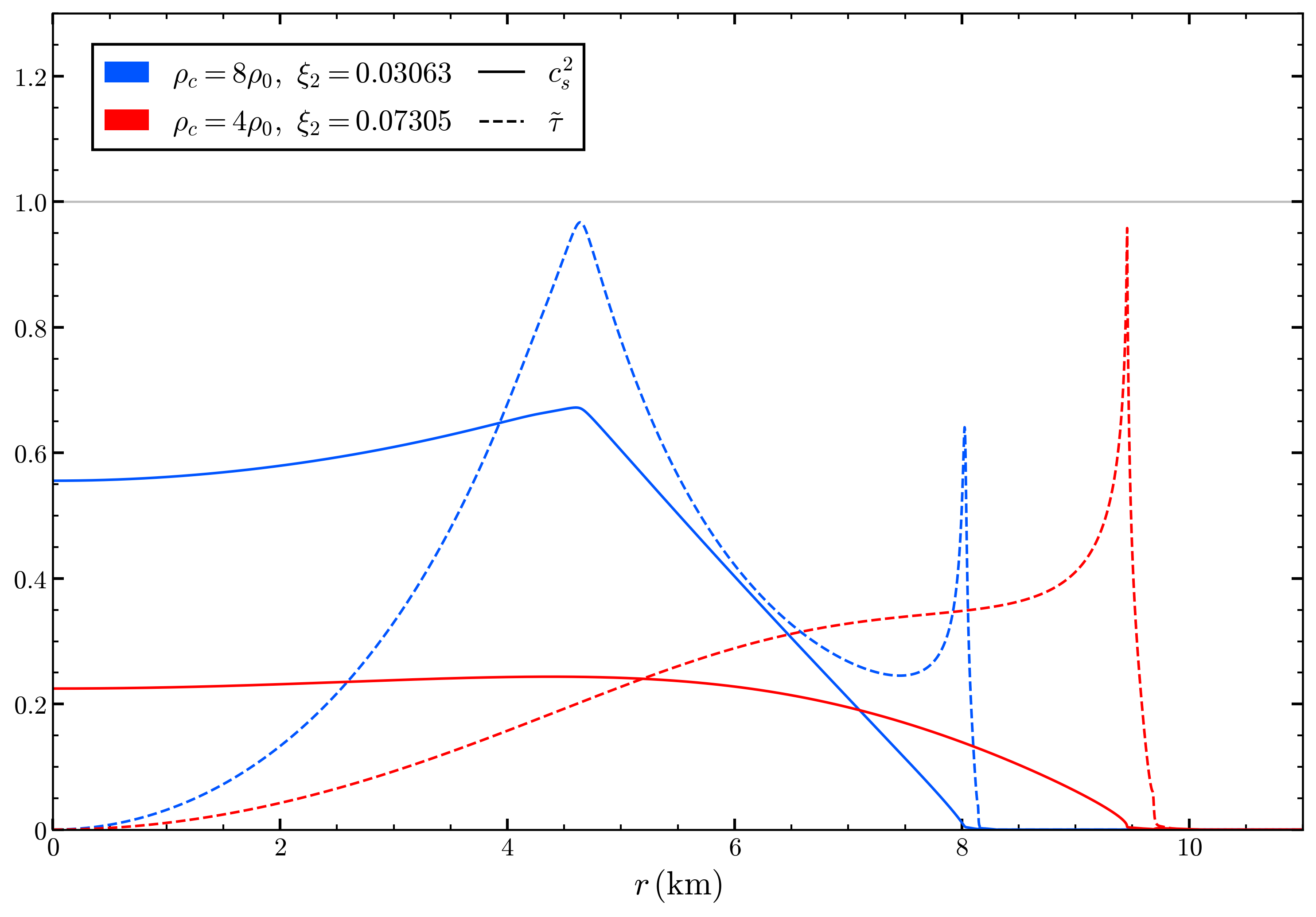}
    \caption{Radial profiles of  $c_s^2$ and $\tau$ for two almost singular configurations. We are using the SLy EoS and  $\lambda= 10$ km. }
    \label{pABrP}
\end{figure}

\begin{figure}[h]
    \centering
    \includegraphics[width=0.54\linewidth]{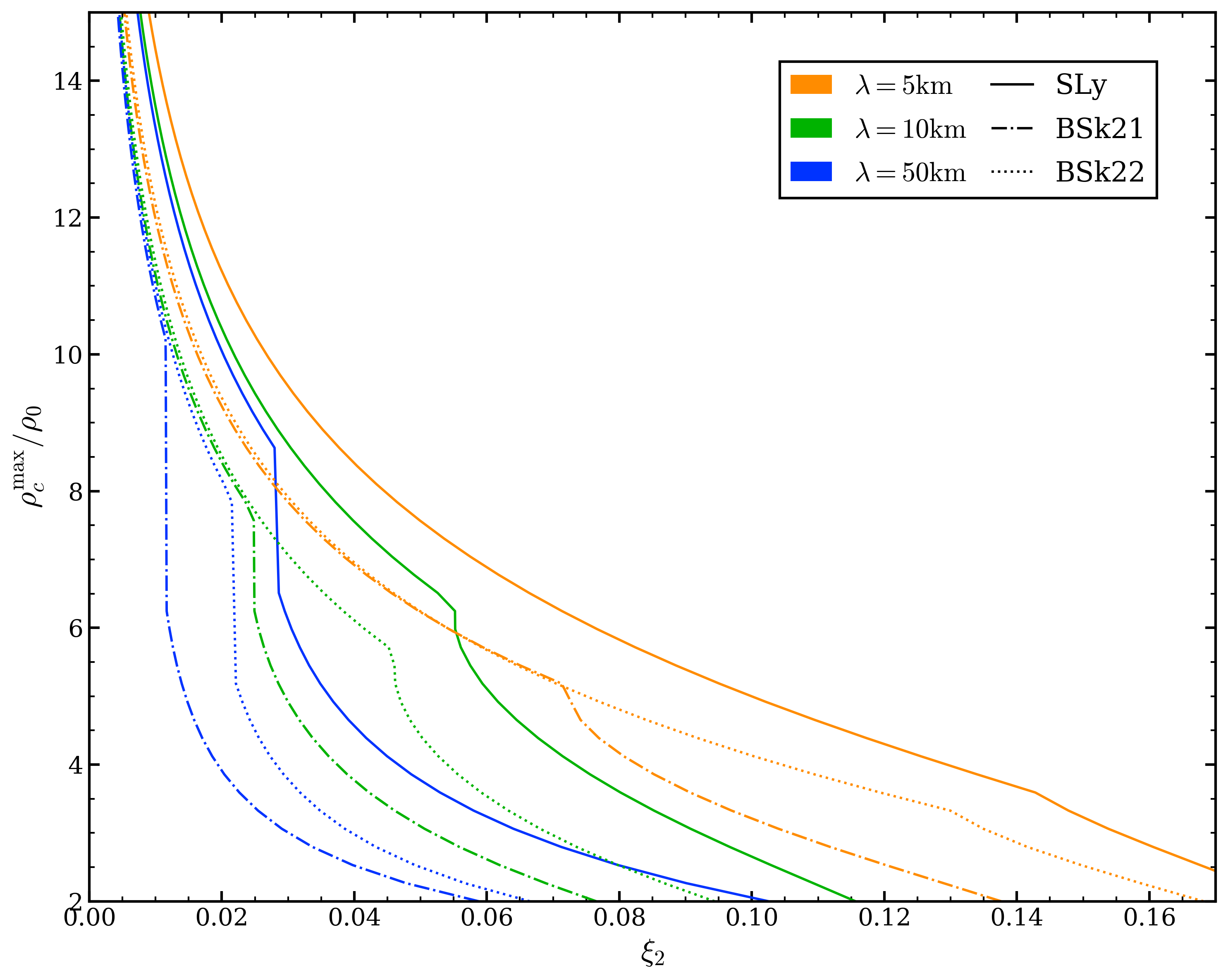}
    \caption{Maximal central density $\rho_c^{\rm max}$ allowed for each $\xi_2$ and for $\lambda=2,5, 10$ and $50$ km, in the case of the SLy EoS. }
    \label{pxirhomax}
\end{figure}

As we stressed in Subsection \ref{subsection_singularity}, the differential system governing the radial profile of a relativistic star breaks down when the quantity $\tilde\tau$, defined in \eqref{b01}, reaches the value $1$.
This condition can be fulfilled either if $\tau$ is large enough or if $c_s$ becomes small enough. This is illustrated in Fig.~\ref{pABrP}, for the SLy EoS, with two distinct values of $\xi_2$ and of $\rho_c$, both near the critical limit associated with a singular system. For the higher value of $\xi_2$ (red curves), 
there is a single peak for $\tilde\tau$, almost reaching $1$,  due to a very small value of $c_s$. By contrast, for the smaller value of $\xi_2$ (blue curves), one observes two peaks: the higher peak corresponds to a large value of $\tau$, whereas the slightly lower peak is associated with a suppressed $c_s$. In such cases, a slight change of the central density can abruptly shift the singularity  from one caused by a  large $\tau$ to one arising  from a small $c_s$.  This explains the sudden slope changes  observed in Fig.~\ref{pxirhomax}, in contrast to Fig.~\ref{pxirhomax_poly} (where $c_s$ never becomes small enough). The precise value where this transition occurs depends on the equation of state and on $\lambda$. 

\begin{figure}[H]
    \centering
    \includegraphics[width=1\linewidth]{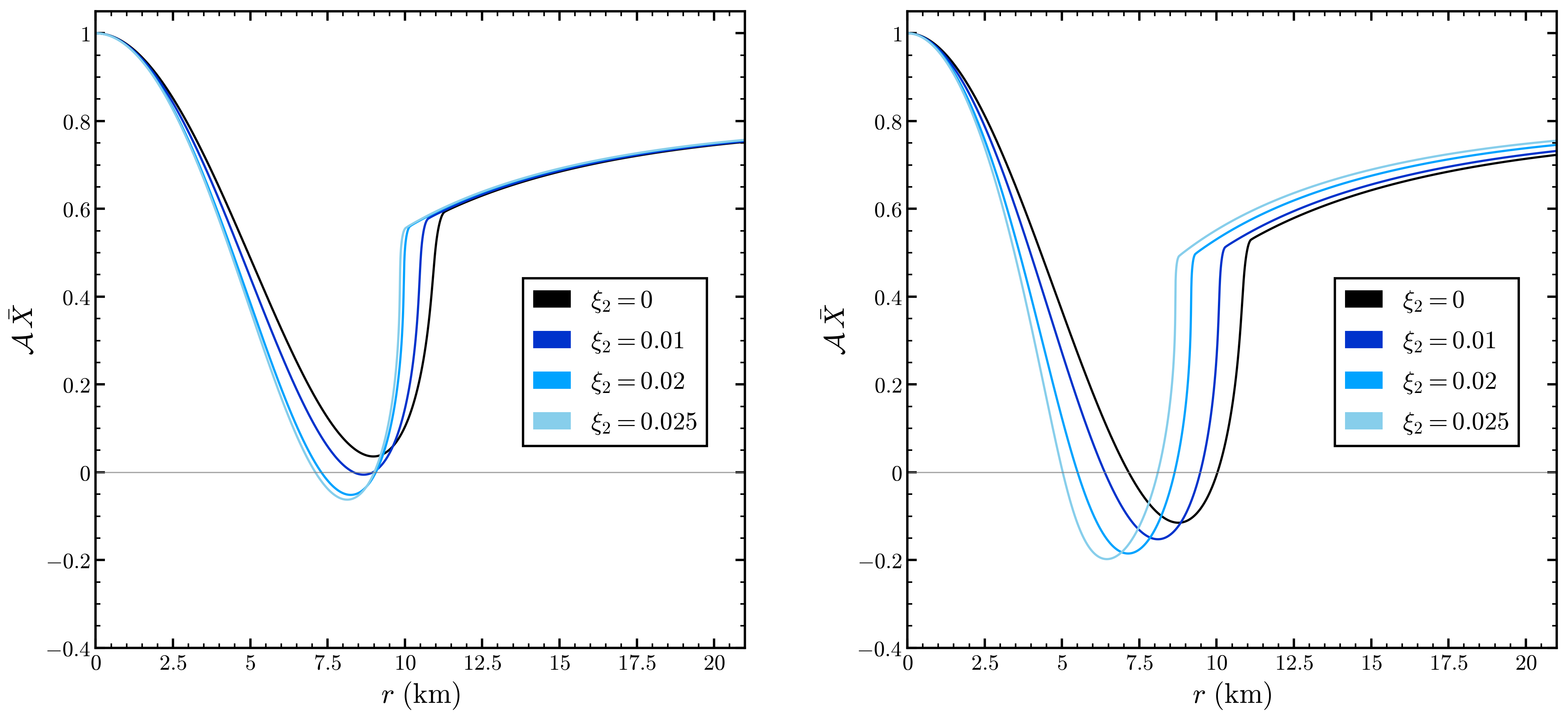}
    \caption{Radial profile of $\cA\,\X$ showing the possible change of sign of the scalar density $X$ in the star interior. Here we consider the SLy equation of state,  two central densities, $\rho_c=6.5 \rho_0$ (left panel) and $\rho_c=8 \rho_0$ (right panel),  $\lambda=50$ km and $\xi_2=0, 0.01, 0.02$ and $0.025$.}
    \label{pX}
\end{figure}

The sign of $X$ outside the star is positive by construction, 
since the gradient of the scalar field is time-like. For most configurations, $X$ is also positive throughout the star. However, in some cases the integration of  the differential system from the center of the star yields a negative value of $X$ for some intermediate range of the radial coordinate, as illustrated in Fig.~\ref{pX}. This boils down to a change in the sign of $\Theta$ in \eqref{Xbar} which in turn depends on the 
matter back-reaction on the scalar kinetic term $X$. This is a very intriguing feature as it means that the gradient of the scalar field 
--- always time-like in the star's core (see Eq.~\eqref{Xc})---  can become space-like in an intermediate region before reverting to  time-like  in the outer layers and exterior. 
We stress that  the differential system  remains regular as $X$ crosses zero. However, it is natural to wonder whether this sign flip could trigger physical instabilities. This would require analyzing  the linear perturbations around such a sign-flipping configuration.

\subsection{Mass-radius relation}
As noticed in the various star configurations shown earlier, the star's radius  in modified gravity differs from that of a GR star with the same central density. 
The mass of the star is also affected. To discuss how the latter depends on the various parameters, it is instructive to go back to the exact exterior solution \eqref{solAq0}. The constant $\mu$ that appears in this equation can be determined numerically from the metric component of the internal solution  evaluated at the  star radius $R$: 
\begin{equation}
    \mu=\frac{R}{2}\left[1-\cA(R)\right]-\xi_\nth \lambda \,\Xi_\nth(R/\lambda)\,.
\end{equation}
According to \eqref{ADM},  the ADM mass is then given by
\begin{equation}
    M\equiv \frac{R}{2}\left[1-\cA(R)\right]+\xi_\nth \lambda\left[\frac{\sqrt{\pi}\,\Gamma(\nth-\frac{3}{2})}{4\, \Gamma(\nth)}- \,\Xi_\nth(R/\lambda)\right]\,,
\end{equation}
which, in the particular case $p=2$, reduces to 
\begin{equation}
    M\equiv \frac{R}{2}\left[1-\cA(R)\right]+\frac{1}{2}\xi_2 \lambda\left[\arctan(\lambda/R)+\frac{R/\lambda}{1+(R/\lambda)^2} \right]\,.
\end{equation}
If the star radius is much larger than $\lambda$,  the term proportional to $\xi_2$ becomes negligible and the mass is given by
\begin{equation}
\label{mass_lambda_small}
    M\simeq \frac{R}{2}\left[1-\cA(R)\right]\qquad {\rm for }\quad R\gg \lambda\,,
\end{equation}
whereas,  in the opposite limit,  we find
\begin{equation}
\label{mass_lambda_large}
    M\simeq \frac{R}{2}\left[1-\cA(R)\right]+\frac{\pi}{4}\xi_2 \lambda\qquad {\rm for }\quad R\ll \lambda\,.
\end{equation}
This latter expression 
applies to the case $\lambda=50$ km plotted in Fig.~\ref{prmass}. Moreover, if  $R\ll \xi_2\lambda$, one even finds that the ADM mass, measured at spatial infinity, is essentially given by the second term, i.e. $M\simeq \pi\xi_2 \lambda/4$.
Note, however, that observers situated at a sufficiently small distance from the star will measure a mass that differs from the ADM mass.

\begin{figure}[h!]
    \centering
    \begin{minipage}{0.48\textwidth}
        \centering
        \includegraphics[width=\linewidth]{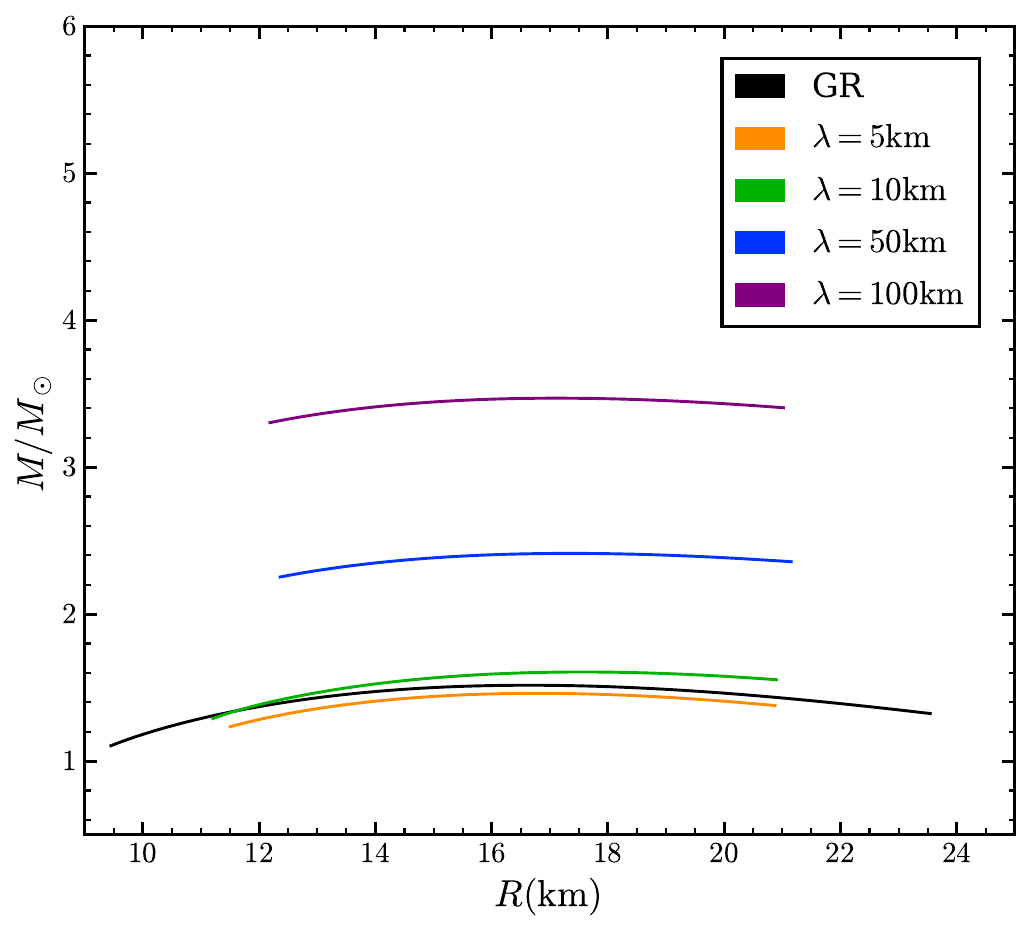}
    \end{minipage}\hfill
    \begin{minipage}{0.48\textwidth}
        \centering
        \includegraphics[width=\linewidth]{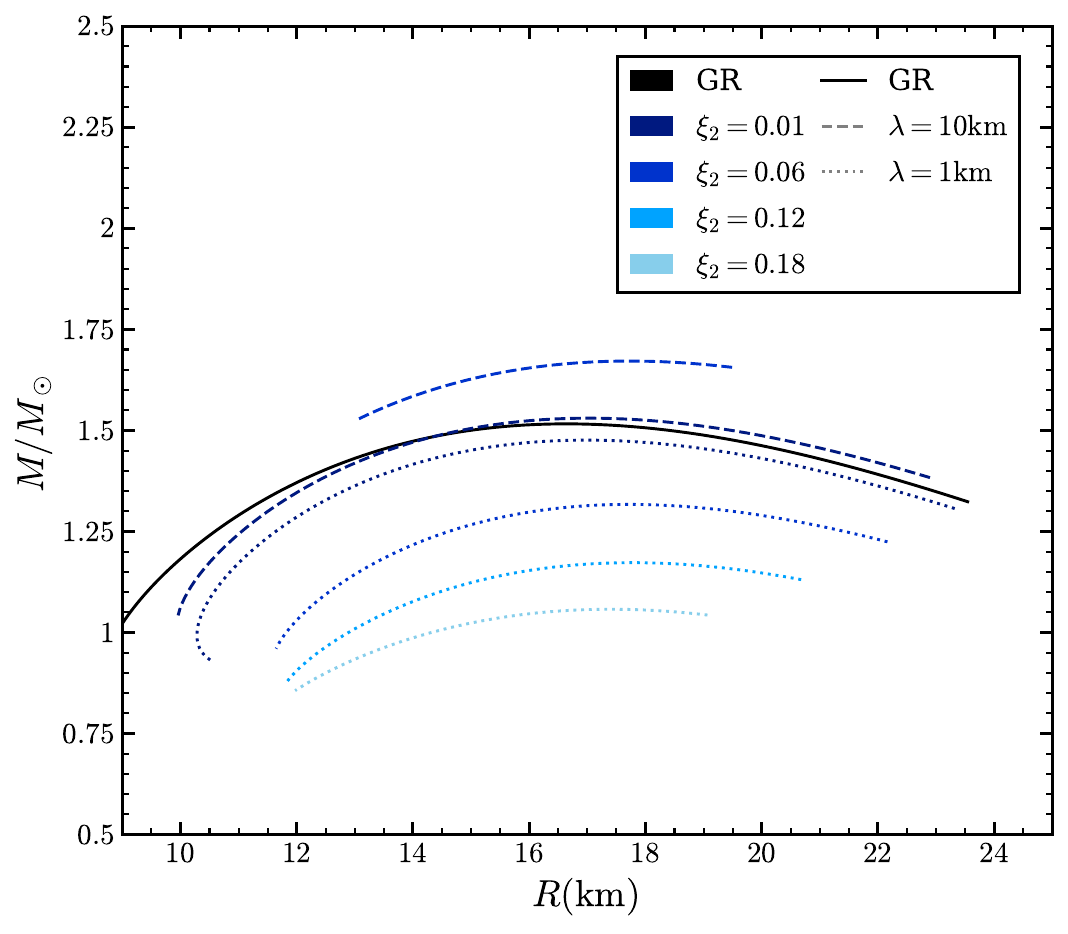}
    \end{minipage}
    \caption{Mass-radius relation $M(R)$ for compact stars with the polytropic EoS  (with $\rho_c \geq 2\rho_0$) . The left panel focuses on the influence of the theory-fixed parameter $\lambda$, for  $\lambda=5, 10, 50$ and $100$ km, with $\xi_2 =0.04$. The right panel focuses on the influence of the star-dependent parameter $\xi_2$, showing the cases $\xi_2 =0.01, 0.6, 0.12$,  for the theory parametrised by $\lambda=10$ km.}
    \label{prm} 
\end{figure}

By varying the central density, we can compute  many neutron star configurations and thus determine the mass-radius relation, depending on the equation of state and on the modified gravity parameters. In Fig.~\ref{prm},  we plot this mass-radius relation in the case of our polytropic EoS.
The GR case, corresponding to  $\xi_2=0$, is plotted as a solid black curve. Overall,  the mass-radius relations exhibit the usual shape expected for equilibrium compact stars: as the central density is increased, the mass  increases, reaches a maximum, and then decreases, corresponding to the threshold of dynamical instability, as determined by the  condition $dM/d\rho_c = 0$.  The colored curves associated with different values  of $\lambda$ or  $\xi_2$ show that this qualitative behavior is preserved, but the curves  are  shifted relative to GR.
In particular, the left  panel shows how the mass-radius relation is affected by the variation of $\lambda$, while $\xi_2$ remains fixed. One observes that for a given radius,  the mass tends to decrease, with respect to GR (black curve), for small values of $\lambda$. By contrast, it increases for large values of $\lambda$. This observation can be explained by the two limiting expressions obtained above. When $\lambda$ becomes large, the mass is dominated by the second term in  \eqref{mass_lambda_large}, which is proportional to $\xi_2 \lambda$. By contrast, for a small $\lambda$, we find that the mass is smaller, at fixed radius. 
The right panel shows how this mass-radius relation changes when $\xi_2$ varies. Essentially, as $\xi_2$ increases, the corresponding curve is further away from the GR curve, in either direction depending on the value of $\lambda$.  
Let us note that the curves stop on the left hand side when the maximal central density is reached. On the right hand side of the curve (see for instance the right panel), the limit corresponds to the minimum central density explored numerically, here $\rho_c= 2\rho_0$.

\begin{figure}[h!]
    \centering
    \includegraphics[scale=0.65]{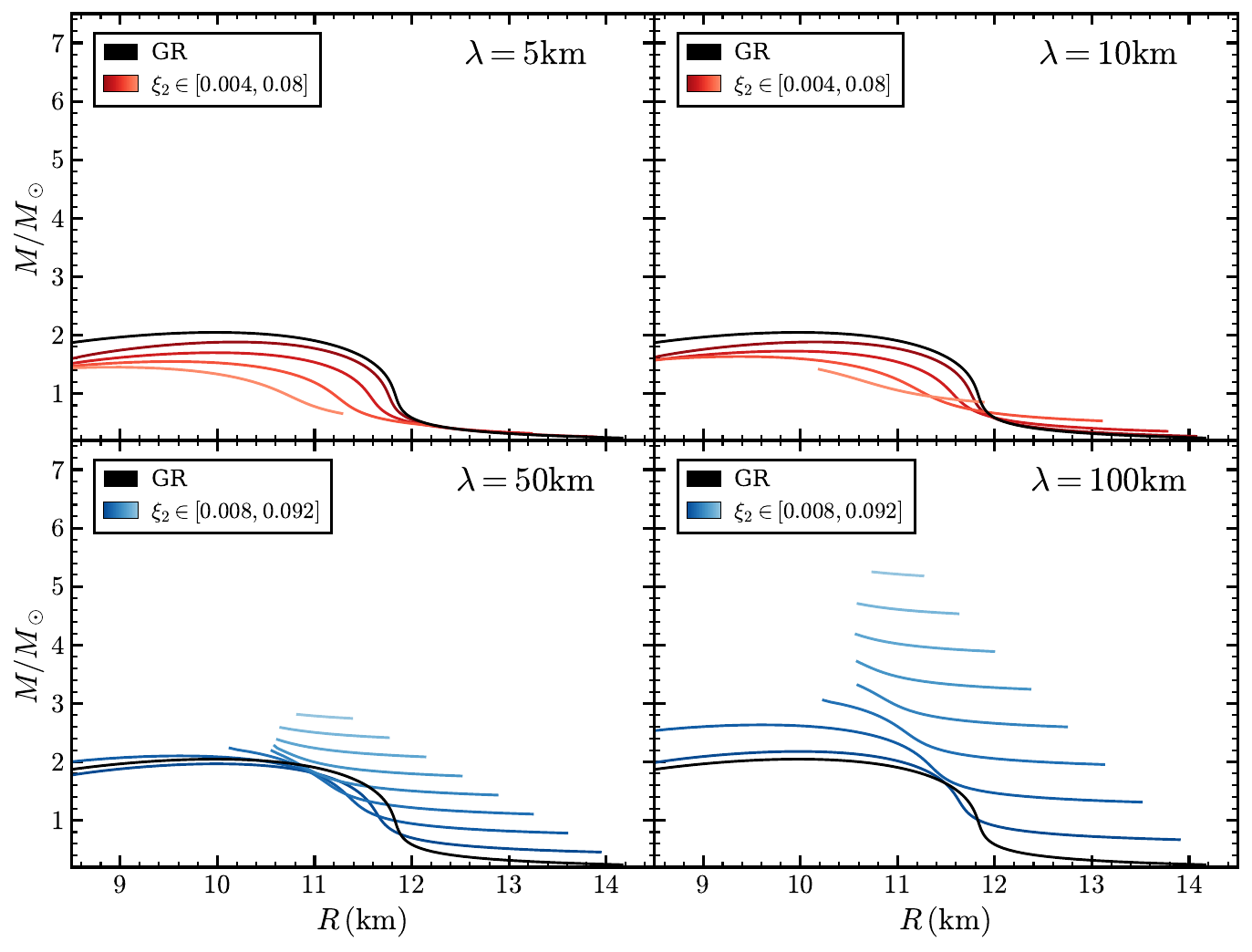}
    \caption{Mass-radius relation $M(R)$ for the SLy EoS. In the top panels we consider $\lambda=5$ and $10$ km respectively for values $\xi_2 =0.004, 0.016, 0.04$ and $0.08$. We notice that for small values of $\lambda$ the mass range (in shades of red) is relatively independent of the value of $\xi_2$. For the bottom panels however (considering the larger values  $\lambda=50$ and $100$ km), the mass range (in shades of blue) differs significantly depending on $\xi_2$, taken here in the array starting from $0.008$ up to $0.092$, with an increment of $0.012$. The curves stop on the left when they reach the maximum central density to avoid a singularity and, on the right, for the minimal central density $\rho_c=2\rho_0$ considered in our computations.}
\label{MRly}
\end{figure}

\begin{figure}[H]
    \centering
\includegraphics[scale=0.6]{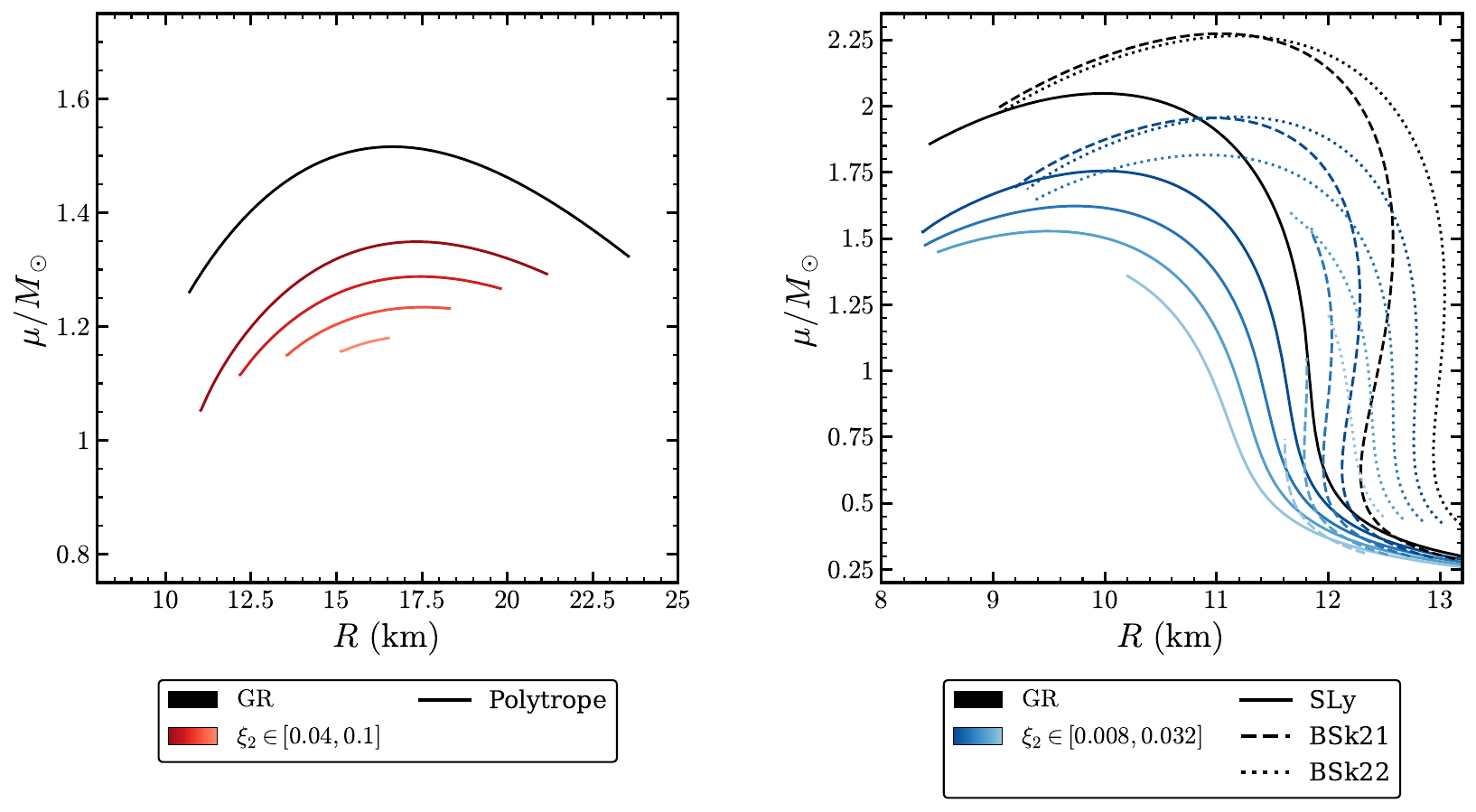}
 \caption{Relation $\mu(R)$ for $\lambda=50$ km, for all EoS studied. The left plot corresponds to the polytropic EoS, for $\xi_2$ from  $0.04$ to  $0.10$ with an increment of $0.02$. The right plot corresponds to the SLy, BSk21 and BSk22 EoS, for $\xi_2$ from  $0.008$ to  $0.032$ with an increment of $0.008$.}
 \label{muRsly}
\end{figure}

Similar plots of the mass-radius relation can be drawn for more realistic equations of state, for example SLy in Fig.~\ref{MRly}. Again, the mass-radius curve is shifted to smaller masses when $\lambda$ is smaller than the star's radius, whereas for large $\lambda$, the second term in \eqref{mass_lambda_large} becomes more and more prominent. 
For this reason, it is instructive to plot the relation between the radius and the mass $\mu$, which is independent of the extra term proportional to $\lambda$. In Fig.~\ref{muRsly}, one clearly sees that the intrinsic mass $\mu$ is reduced with respect to the mass of a GR star with the same  radius. We also note that, for realistic equations of state, the mass $\mu$ is more sensitive to the parameter $\xi_2$ than for polytropic equations of state (compare the increment of $\xi_2$ in the left and right plots of Fig.~\ref{prm}). This can be related to the property that realistic EoS have a higher adiabatic index at high  density, which enhances the impact of the modified gravity effects on the star profile.

\subsection{Case of negative scalar charge}
\begin{figure}[h!]
    \centering
    \includegraphics[scale=0.5]{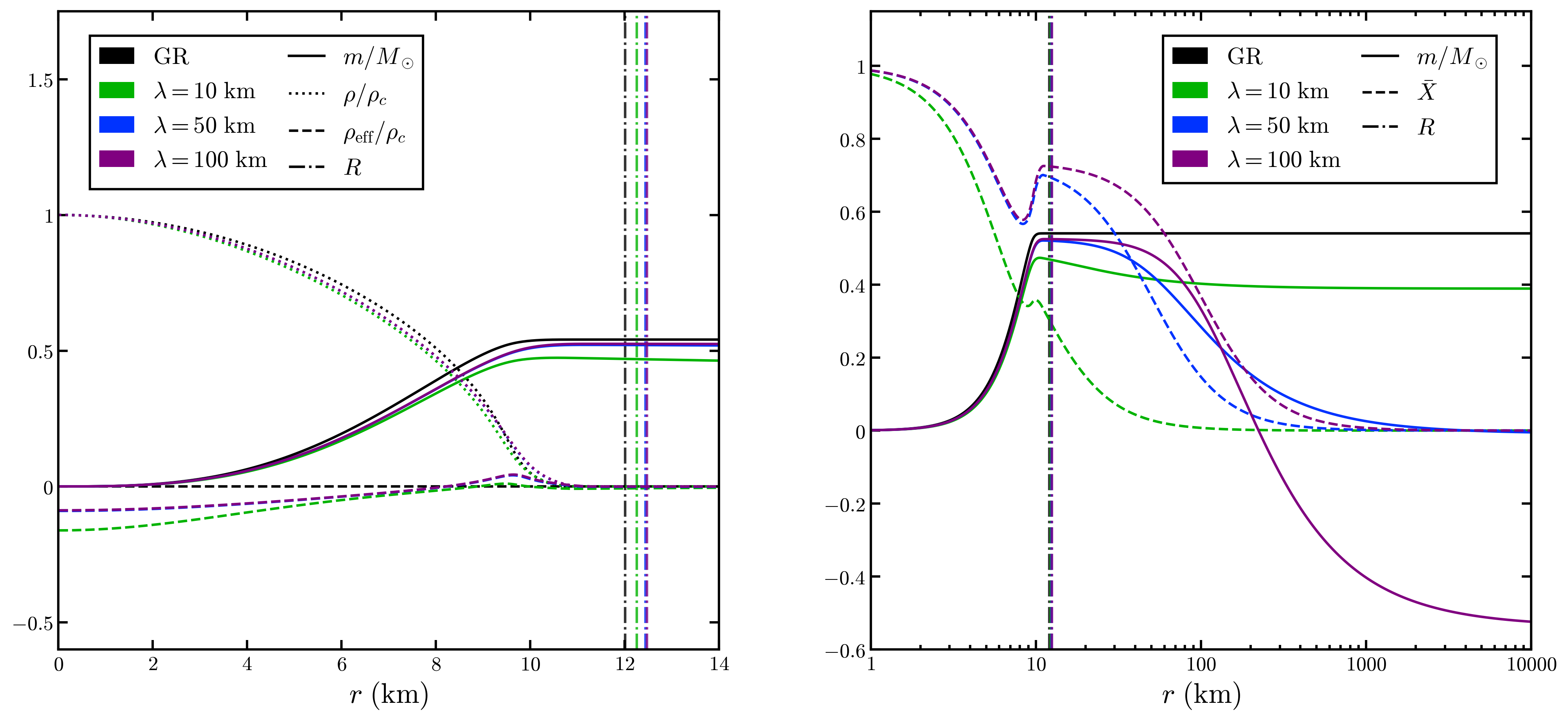}
    \caption{Radial profiles of $m$, $\rho$, $\rho_{\rm eff}$ and $\X$ for the SLy EoS, for $\lambda=10, 50$ and $100$ km, $\rho_c=3\rho_0$ and and $\xi_2=-0.02$. The left plot gives the profile of $m$, $\rho$ and $\rho_{\rm eff}$ within the  star, while the right one focuses on the behavior of $m$ and $\X$ at larger distances. Note that for $\lambda=100$ km (purple curve), $m$ eventually becomes negative.}
    \label{prmassn} 
\end{figure}

\begin{figure}[h!]
    \centering
\includegraphics[width=\textwidth]{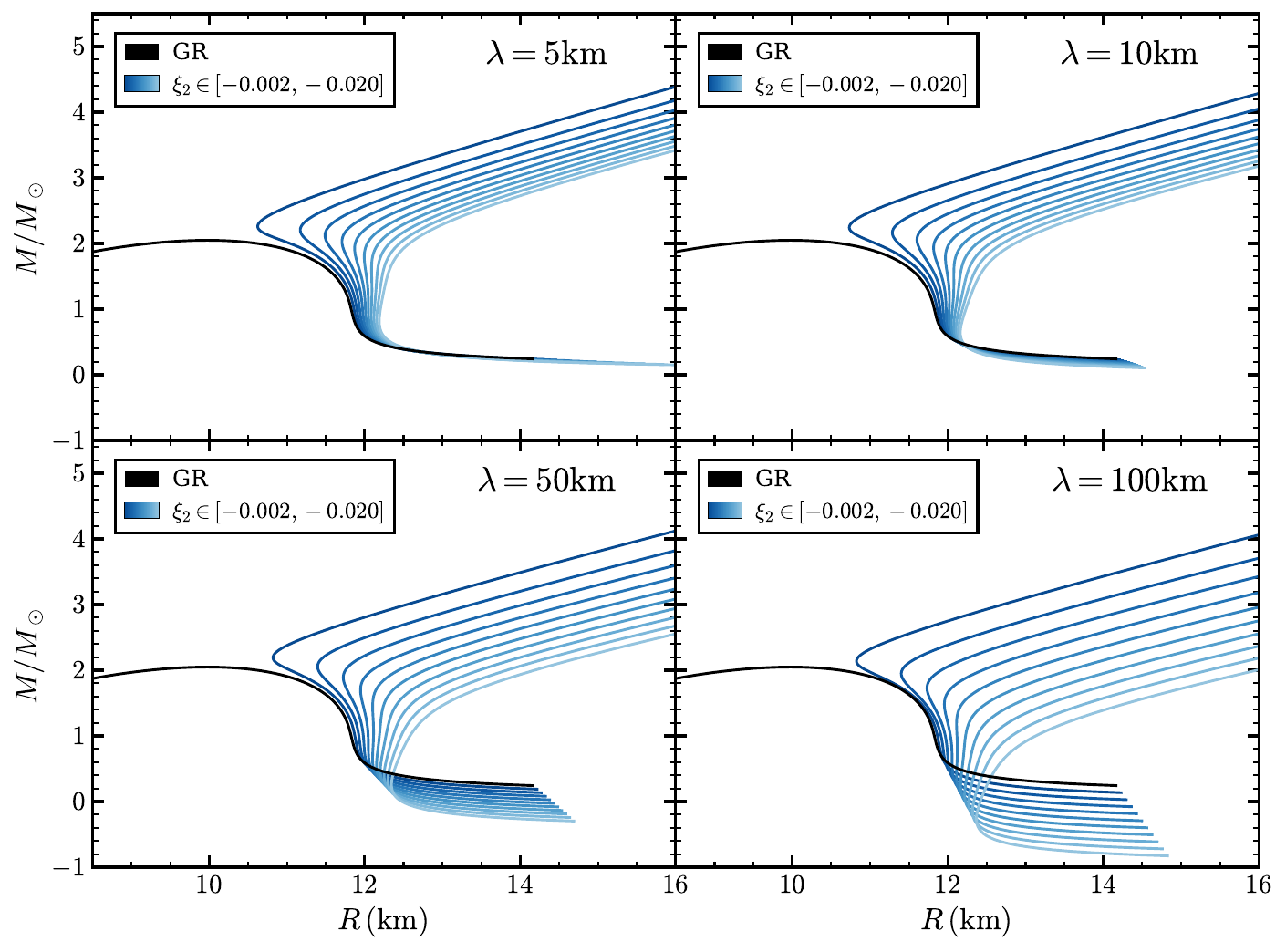}
    \caption{Mass-radius relation $M(R)$ for the SLy EoS in the case of a negative scalar charge, for $\lambda=5, 10, 50$ and $100$ km and $\xi_2$ varying from  $-0.002$ to $-0.020$ with an increment of $0.002$.}
\label{mRslynegative}
\end{figure}

We now  discuss briefly the cases where the coupling parameter $\xi_\nth$ takes negative values, which appears to  lead to puzzling --- and likely problematic --- consequences. 
For $\xi_\nth<0$,
the effective energy density $\rho_{\rm eff}$ is now negative in the innermost part of the star, and positive near the surface (because the dominant term $\X'/\X$ in Eq.\eqref{rho_eff_bis}  becomes positive near the surface, as shown
in Fig.~\ref{prmassn}, while the effective pressure is positive, acting against the gravitational 
pull
in addition to the ordinary pressure. Therefore,  neutron stars now tend to be less compact, leading to larger radii for a given mass. This behavior can be  observed in the mass–radius diagram of Fig.~\ref{mRslynegative}, showing a larger radius compared to the GR prediction.

A surprising consequence of a negative effective energy density is that the ADM mass can become negative.
Indeed, by integrating \eqref{m'}, one finds
\begin{equation}
    M=M_{\rm m}+M_{\rm eff}\,
\end{equation}
with 
\begin{equation}
    M_{\rm m}= 4\,\pi\int_0^R\rho\,r^2 dr\,, \qquad M_{\rm eff}=M^{\rm in}_{\rm eff}+M^{\rm out}_{\rm eff}\equiv 4\,\pi \int_0^R\rho_{\rm eff}\,r^2 dr+ 4\,\pi \int_R^\infty\rho_{\rm eff}\,r^2 dr \,,
\end{equation}
where we have distinguished the interior and exterior effective mass contributions\footnote{One can even find  cases where $|M^{\rm in}_{\rm eff}|>M_{\rm m}$,  for low values of $\lambda$ and $\rho_c$, along with a sufficiently  high value of $|\xi_2|$.}.

The matter contribution to the mass, $M_{\rm m}$, is of course positive, but the interior effective mass $M^{\rm in}_{\rm eff}$ is negative. Moreover, the external effective mass is  given by 
\begin{equation}
   M^{\rm out}_{\rm eff} = \xi_\nth\left[
    \frac{\sqrt{\pi}\,\Gamma(\nth-\frac32)}{4\, \Gamma(\nth)}-\Xi_\nth(R/\lambda)\right] \lambda\,,
\end{equation}
where the term in the brackets is always positive. Thus $M^{\rm out}_{\rm eff}$ is also negative for negative $\xi_\nth$ and, when $\lambda \gg R$, becomes proportional to $\lambda$:
\begin{equation}
   M^{\rm out}_{\rm eff} \simeq  \xi_\nth
    \frac{\sqrt{\pi}\,\Gamma(\nth-\frac32)}{4\, \Gamma(\nth)}\, \lambda  \qquad {\rm for}\quad \lambda\gg R\,.
\end{equation}
It is thus possible to obtain stars with arbitrarily large negative mass by increasing $\lambda$.
This can be seen in Fig.~\ref{mRslynegative}, for  $\lambda=50$ km and $\lambda=100$ km, with small  central energy density.

At very high central densities, the  star exhibits  unusual behaviour: both its mass and radius increase with $\rho_c$. Indeed, as $\rho_c$ rises,  the effective energy density in the core becomes increasingly negative, while  the effective pressure grows simultaneously. This causes ordinary matter to rarefy in the core and accumulate in the outer layers. The additional pressure from the effective fluid both supports a higher mass and, through its repulsive effect, increases the radius.

\newpage
\section{Conclusions}
\label{section_conclusion}

We have constructed neutron star solutions within a subfamily of DHOST theories, exploiting their global shift and parity symmetries. These theories are  parametrised by two coupling constants and a positive  exponent $p$, which we  fixed to $2$ to obtain a simple closed-form solution for the vacuum spacetime.  

As in GR, the neutron star interior is modeled as a perfect fluid, with the solution matching an exact vacuum solution  at the star's surface. Unlike previous solutions in DHOST theories, the neutron stars we have constructed  carry a finite scalar charge $\xi_\nth$ (defined in Eq.~\eqref{xi}), which acts as  a free parameter, analogous to the  mass or electric/magnetic charge in GR. In addition, the solutions depend on a characteristic length scale $\lambda$ --- fixed by the theory --- which sets the range of the scalar field's effects on the geometry. The exterior solution corresponds to the   black holes  with primary scalar hair recently discovered in~\cite{Bakopoulos:2023fmv} and further extended in the literature (see for example \cite{Bakopoulos:2023sdm}, \cite{Baake:2023zsq},\cite{Charmousis:2025xug}).

For sufficiently high values of $\xi_\nth$, the neutron star solutions we obtained exhibit striking features that depart significantly from their GR counterparts (with the same central energy density and equation of state). The origin of these differences can be traced back to the scalar field kinetic term,  which now depends on  the  matter (see Eq.~\eqref{Xbar}) and can drastically change what happens in the interior of the compact object by a cascading effect in the field equations.{\footnote{Note that although the scalar field is not directly coupled to matter, it is the nonlinear nature of the modified TOV equations  that lead to these backreaction effects in the star's interior.}} As a result, the matter energy density (and pressure) can increase when going outwards in  the star's core, in  contrast with the usual GR behaviour. 
One may suspect that this feature is associated with  instabilities\footnote{It has been noted that some matter instabilities may emerge in certain classes of DHOST theories (see for example \cite{Babichev:2018rfj, Saltas:2019ius}).} and deserves further investigation, particularly of radial perturbations, which we plan to do study in future work. Furthermore, we have found that, at a finite distance inside the star, a singularity of the field equations may appear for a sufficiently large (positive) scalar $\xi_\nth$ or a small $\lambda$, when the central density is high. Another surprising property of the configurations we have obtained is the possibility for the scalar field gradient to evolve from time-like at the star's center to space-like inside an intermediate region of the star to time-like again near the surface and outside. 

As mentioned earlier, an interesting property of our system is the very particular back-reaction of matter on the scalar field kinetic term $X$,  as  depicted by Eq.~\eqref{Xbar} in the star's interior. Reciprocally, the impact of the scalar field on the star profile is embodied in the radially-dependent  effective energy density $\rho_{\rm eff}$ and pressure $\p_{\rm eff}$, with equation of state $\p_{\rm eff}=-\rho_{\rm eff}$. Interestingly, a jump in the energy density and pressure due to a phase transition inside the neutron star (see for example \cite{Ventagli:2024cho}) would have no direct impact on the quantities $\Theta$, defined in \eqref{Xbar}, and $\bb$, defined in \eqref{b0}, that characterise the beyond-GR effects since they depend on the combination $\rho+P$. 

Quite recently, black holes with primary hair were found \cite{Charmousis:2025jpx, Fernandes:2026rjs,Charmousis:2026dbi} within a generic class of vector-tensor theories (see for example \cite{Heisenberg:2017mzp}). These acquired a quite distinct construction from that used in \cite{Bakopoulos:2023fmv} which is interestingly related to a Weyl rather  than Levi-Civita  connection in higher dimensional Lovelock theory \cite{BeltranJimenez:2014iie,Bahamonde:2025qtc}. It was found \cite{Charmousis:2026dbi} that one can have an arbitrarily small cosmological constant as primary hair. It would be worthwhile to investigate the effects of such a parameter on compact objects, like neutron stars, within these theories and see what it may entail in terms of observations.

We have seen that, \emph{outside} a compact object, 
the scalar field interferes
in the local versus far-away gravitational pull of the object, with respect to standard GR. It would be interesting to study whether there exists some astrophysical system where such an effect could yield valuable constraints on $\lambda$ and $\xi_p$. 

To determine whether the  strange behaviours  we have observed — at least for some values of the parameters — such as the $X=0$ crossing, the non-monotonic matter profile, are physically relevant, it would  be crucial to consider linear perturbations of our stars.  This  would presumably show that some configurations are unstable, thereby imposing more stringent constraints on our beyond-GR parameters.

Another interesting direction would be to extend our study  to solutions associated with non-homogeneous exterior metrics, obtained via disformal transformations of the neutron star solutions at hand. In particular, it would be valuable to investigate whether the matter singularities observed here  could be alleviated under  certain of disformal deformations.

\subsection*{Acknowledgements}
HB would like to sincerely thank both laboratories  \textit{Astroparticule et Cosmologie} (APC) and \textit{Laboratoire de Physique des 2 Infinis Irène Joliot Curie} (IJCLab) for hosting him during  the initiation of this work. CC would like to thank fruitful discussions with George Pappas during his visits in the Observatory of the University of Thessaloniki.

\newpage
\appendix

\section{Expansion near the center of the equations of motion}\label{appB}
In this appendix, we present the behavior of the functions $\cA$, $\cB$ and $P$ near the center of the star.  
Regularity at the center imposes the  boundary conditions are $\cA'(0)=\cB'(0)=\p'(0)=0$ and $\cB(0)=1$, which leads to the following expansions  near $r=0$:
\begin{eqnarray}
    &\cA(r)=\cA_c+\cA_2 \,r^2+\dots\\
    &\cB(r)=1+ \cB_2 \,r^2+\dots\\
    & \p(r)=\p_c+\p_2 \,r^2+\dots,
\end{eqnarray}
where $P_2$, $\cA_2$, $\cB_2$ and $\p_2$ are constants.  To determine  these constants  in terms of $\rho_c$ and $\p_c$, we substitute the above  expressions into the background equations (\ref{eA}), (\ref{eB}) and (\ref{eP}) and obtain
\begin{eqnarray}
    \cB&=&1- \left(\frac{\kappa  }{3}\rho _c+\frac{ \xi _\nth (2 \nth-1)}{6 \cA_c^{\nth}\lambda ^2 \nth}\left(3 \kappa  \lambda ^2  (\rho _c+3 \p_c)+\frac{4 \nth}{2 \nth-1}\right)\right)r^2,\label{hc}\\
    \cA/\cA_c &=&1+   \left(\frac{\kappa }{6}  \left(3 \p_c+\rho _c\right)-\frac{ \xi _\nth (2 \nth-1) }{6 \cA_c^{\nth} \lambda ^2 \nth}\left(3 \kappa  \lambda ^2  (\rho _c+ \p_c)+\frac{4 \nth}{2 \nth-1}\right)\right)r^2,\label{fc}\\
    \p&=&\p_c -\frac{\p_c+\rho _c}{2}\left(\frac{\kappa }{6}   \left(3 \p_c+\rho _c\right)-\frac{ \xi _\nth (2 \nth-1)}{6 \cA_c^{\nth}\lambda ^2 \nth} \left(3 \kappa  \lambda ^2  (\rho _c+  \p_c)+\frac{4 \nth}{2 \nth-1}\right)\right)r^2\,.
    \label{Pc}
\end{eqnarray}
As a consequence, the kinetic term near the center is given by 
\begin{eqnarray}
\label{Xc}
  X=  \frac{q^2}{2 \cA_c}\left(1- \left(\frac{\kappa  (3 \nth-1) \left(\p_c+\rho _c\right)}{2 \nth}+\frac{1}{\lambda ^2}\right)r^2\right)+\dots
\end{eqnarray}

For $\xi_\nth>0$, the local mass $m=(1-\cB)/2$ is always positive.  By contrast, for sufficiently large $\xi_\nth$ the sign of $P'(0)$ (respectively $\cA'(0)$) can become positive (respectively negative).
Imposing that the pressure, like in GR, decreases from the center, i.e. $P'(0)<0$, yields an upper bound on the parameter $\xi_\nth$, which can be expressed as
\begin{eqnarray}
    \frac{\xi _\nth}{\mathcal{A}_c^\nth}<\frac{\kappa  \lambda ^2 \nth  \left(3 
    \p_c+\rho _c\right)}{3 \kappa  \lambda ^2 (2 \nth-1) \left(\p_c+\rho _c\right)+4 \nth}\,.
\end{eqnarray}

 Finally, from Eq.~(\ref{rho_eff_bis}) and from the expansions of $\cA$, $\cB$ and $P$, one can evaluate $\rho_{\rm eff}$ at the center of the star,
\begin{equation}
\label{rho_eff_bis_per}
    \kappa \rho_{{\rm eff},c}\equiv      \frac{\xi_\nth }{\cA_c^p} \left(\frac{\kappa  (6 \nth-1) \left(\p_c+\rho _c\right)}{2 \nth}+\frac{2}{\lambda ^2}\right)\,,
\end{equation}
where the sign of $\rho_{\rm eff}$ depends on the scalar charge $\xi_\nth$. This agrees with  the numerical results of Section \ref{section_numerics}.
\newpage
\section{Coefficients $b_i$ for the analytic expression of realistic equations of states}
\label{App:tables}
In this section, we give the coefficients  $b_i$  that appear in  the analytic expression \eqref{EOS} for realistic equations of state.
\begin{table}[H]
\centering
\begin{tabular}{||c| c||}
\hline\hline
$b_i$ &  {SLy} \\
\hline\hline
1  & 6.22 \\
2  & 6.121 \\
3  & 0.005925 \\
4  & 0.16326 \\
5  & 6.48 \\
6  & 11.4971 \\
7  & 19.105 \\
8  & 0.8938 \\
9  & 6.54 \\
10 & 11.4950 \\
11 & -22.775 \\
12 & 1.5707 \\
13 & 4.3 \\
14 & 14.08 \\
15 & 27.80 \\
16 & -1.653 \\
17 & 1.50 \\
18 & 14.67 \\
\hline\hline
\end{tabular}
\begin{tabular}{||c  |c |c||}
\hline\hline
$b_i$ &  {BSk21} & {BSk22} \\
\hline\hline
1 &  4.857 & 6.682\\
2 &  6.981 & 5.651\\
3 & 0.00706 & 0.00459 \\
4 &  0.19351 & 0.14359\\
5 &  4.085 & 2.681\\
6 &  12.065 & 11.972\\
7 &  10.521 &  13.993 \\
8 &  1.5905 & 1.2904 \\
9 &  4.104 & 2.665\\
10 &  -28.726 & -27.787 \\
11 &  2.0845 & 2.0140 \\
12 &  4.89 & 4.09\\
13 &  14.302 & 14.135\\
14 &  22.881 & 28.03\\
15 &  -1.7690 & -1.921\\
16 &  0.989 & 1.08 \\
17 &  15.313 &14.89\\
18 &  0.091 & 0.098\\
19 &  4.68 & 4.75\\
20 &  11.65 & 11.67\\
21 &  -0.086 &-0.037 \\
22 &  10.0 & 11.9\\
23 &  14.15 & 14.10\\
\hline\hline
\end{tabular}\caption{Coefficients $b_i$  for SLy (from \cite{Potekhin:2013qqa}) and coefficients $b_i$ for  BSk21 and BSk22 (from \cite{Haensel:2004nu,Pearson:2018tkr}).
\label{table2}
}
\end{table}

\newpage
\bibliographystyle{utphys}
\bibliography{DHOST_NS_biblio1}
\end{document}